\documentclass[11pt,a4paper]{article}
\usepackage[utf8]{inputenc} 
\usepackage[english]{babel}
\usepackage[T1]{fontenc}
\usepackage{graphicx,amssymb,amsmath,wasysym}
\usepackage[bookmarks=false]{hyperref}
\usepackage[nosort]{cite}

\begin{document}
\baselineskip=15.5pt

\thispagestyle{empty}

\begin{flushright}
Bonn-Th-2011-08\\
CFTP/11-007\\
\end{flushright}

\vspace{.5cm}

\begin{center}

{\LARGE\sc{\bf Constraining the Milky Way Dark Matter Density Profile
    \\[1ex]  with Gamma--Rays with {\em Fermi}--LAT}}

\vspace*{9mm}
\setcounter{footnote}{0}
\setcounter{page}{0}
\renewcommand{\thefootnote}{\arabic{footnote}}

\mbox{ {\large \bf Nicolás Bernal$^{1}$ and Sergio Palomares-Ruiz$^2$}}

\vspace*{0.9cm}

{\it $^1$Bethe Center for Theoretical Physics and Physikalisches
  Institut,\\ Universit\"at Bonn, Nu\ss allee 12, D-53115 Bonn,
  Germany}\\
\vskip 0.3cm
{\it $^2$Centro de Física Teórica de Partículas (CFTP),\\
Instituto Superior Técnico, Avenida Rovisco Pais 1, 1049-001 Lisboa,
Portugal}\\ 
\vskip5mm
e-mails: {\tt nicolas@th.physik.uni-bonn.de,
  sergio.palomares.ruiz@ist.utl.pt} 
\vskip 5mm

\end{center}

\vspace{1cm}

\begin{abstract}
We study the abilities of the {\it Fermi}--LAT instrument on board of
the {\it Fermi} mission to simultaneously constrain the Milky Way dark
matter density profile and some dark matter particle properties, as
annihilation cross section, mass and branching ratio into dominant
annihilation channels.  A single dark matter density profile is
commonly assumed to determine the capabilities of gamma-ray
experiments to extract dark matter properties or to set limits on
them.  However, our knowledge of the Milky Way halo is far from
perfect, and thus in general, the obtained results are too optimistic.
Here,  we study the effect these astrophysical uncertainties would
have on the determination of dark matter particle properties and
conversely, we show how gamma-ray searches could also be used to learn
about the structure of the Milky Way halo, as a complementary tool to
other type of observational data that study the gravitational effect
caused by the presence of dark matter.  In addition, we also show how
these results would improve if external information on the
annihilation cross section and on the local dark matter density were
included and compare our results with the predictions from numerical
simulations.
\end{abstract}

\newpage

\tableofcontents

\newpage

\section{Introduction}
\label{introduction}

One of the mysteries of modern physics is the presence of an unknown
non-luminous and non-baryonic matter in the Universe (see
Refs.~\cite{Jungman:1995df, Bergstrom:2000pn, Munoz:2003gx,
  Bertone:2004pz, Bertone:2010} for reviews), the so-called dark
matter (DM).  However, although there are many pieces of evidence in 
favor of the existence of DM and which indicate that it constitutes
about 80\% of the total mass of the Universe~\cite{Tegmark:2006az,
  Komatsu:2008hk, Komatsu:2010fb}, it is worth noting that they each
infer DM's presence uniquely through its gravitational influence. In
other words, we currently have no conclusive evidence for DM's
non-gravitational interactions.  Thus, the origin and most of the
properties of DM (mass, spin, couplings) remain unknown.

Although many particles have been proposed as DM candidates (see,
e.g., Refs.~\cite{Bertone:2004pz, Bergstrom:2009ib} for a
comprehensive list), a weakly interacting massive particle (WIMP),
with mass lying from the GeV to the TeV scale, is one of the most
popular ones.  WIMPs  can arise in extensions of the Standard Model
such as supersymmetry (e.g., Ref.~\cite{Jungman:1995df}), little Higgs
(e.g., Ref.~\cite{Birkedal:2006fz}) or extra-dimensions models (e.g.,
Ref.~\cite{Hooper:2007qk} ) and are usually stable and thermally
produced in the early Universe with an annihilation cross section
(times relative velocity) of $\langle \sigma v \rangle \sim 3 \times
10^{-26}$~cm$^3$/s, which is the standard value that provides the
observed DM relic density.  Thus, WIMPs are cold DM (CDM) candidates.

Within the framework of CDM, as favored by observations, structure
forms hierarchically bottom-up, with DM collapsing first into small
halos, which then accrete matter, merge and eventually give rise to
larger halos~\cite{White:1977jf, Blumenthal:1984bp, White:1991mr}.
Within this paradigm, the properties of luminous galaxies are expected
to be related to those of DM halos~\cite{Mo:1997vb}, which carry with
them gas that cools and collapses to their center to form galaxies. 

Over the past decades, the progress in high-resolution N-body
simulations has permitted a better understanding of the structure of
CDM halos.  It has been shown that density profiles of CDM halos can
be reasonably well described by an universal form, independent of the
halo mass, which predicts cuspy mass distributions.  Although the
first analytical analysis of this kind predicted that the radial
distribution of DM follows a simple power-law~\cite{Gunn:1977}, it was
later shown with N-body simulations that the slope of the DM
distribution is not constant and changes with the distance to the
center of the halo.  It was Navarro, Frenk and White (NFW) who first
proposed a simple two-parameter fitting formula to describe
spherically averaged DM density profiles~\cite{Navarro:1995iw,
  Navarro:1996gj}.  However, further simulations with improved
numerical resolution started to show some systematic
differences~\cite{Burkert:1995yz, Moore:1999gc, Ghigna:1999sn,
  Fukushige:1996nr, Subramanian:1999bg, Fukushige:2000ar,
  Klypin:2000hk, Taylor:2001bq, Power:2002sw, Ricotti:2002qu,
  Fukushige:2003xc, Diemand:2004wh, Gentile:2004tb, Salucci:2007tm},
mainly in the innermost regions of CDM halos.  Thus, in order to
reproduce several of these features, as the gradual shallowing of the
density profile towards the center of the halo, an improved
three-parameter fitting formula was proposed~\cite{Navarro:2003ew},
the so-called Einasto profile~\cite{Einasto}. 

Nevertheless, these prescriptions only represent the mean of all
simulated halos for a given mass at a given redshift. The scatter
with respect to these mean values~\cite{Jing:1999km, Bullock:1999he,
  Wechsler:2001cs, AvilaReese:2005fm, Maccio':2006nu, Maccio':2008xb}
is thought to be related to the different halo formation histories and
to the evolution of the expanding Universe~\cite{Navarro:1996gj,
  Eke:2000av, Zhao:2002rm, Zhao:2003jf, Neto:2007vq, Gao:2007gh,
  Duffy:2008pz, Zhao:2008wd, Klypin:2010qw, MunozCuartas:2010ig}.
Thus, these average values do not necessarily describe accurately the
DM halo of our own Galaxy.  Indeed, it is not clear that the Milky Way
is a prototypical spiral galaxy.  For instance, while the presence of
the two Magellanic Clouds points to a more massive halo, this would also
imply higher chances for a recent merger that could have destroyed the
thin galactic disc~\cite{BoylanKolchin:2009an}.

However, probing the inner structure of the Milky Way DM halo is a
very challenging task and current data do not allow to distinguish
among different density profiles~\cite{Catena:2009mf, Weber:2009pt}.
This is particularly important for the astrophysical detection of DM, 
as these searches are critically sensitive to the structure of the
Milky Way DM halo.  Direct detection experiments, searching for
signals of nuclear recoil of DM scattering off nuclei, depend on the
local DM density and velocity distribution.  On the other hand,
indirect searches look for the products of DM annihilation (or decay),
which include antimatter, neutrinos and photons.  In the case of DM
annihilation, the luminosity depends quadratically on the DM density
and the largest uncertainties come from the region where we expect the
highest (neutrino or gamma-ray) signal, i.e., the galactic center
(GC).  Hence, a better understanding of the Milky Way halo is required
to determine the potential reach of current and future experiments.
Conversely, astrophysical searches, and in particular indirect DM
gamma-ray searches, could also be used to learn about the distribution 
of DM in the innermost regions of the Milky Way halo, beyond the limit
of convergence of numerical simulations.  Thus, in case of a positive
signal, these searches could be useful not only to learn about the DM
nature, but also as a complementary tool to constrain the Milky Way DM
density profile, which may not be necessarily representative of the
general characteristics obtained in numerical simulations.

In this work, we study the abilities of the Large Area Telescope ({\it
  Fermi}--LAT) instrument on board of the {\it Fermi Gamma-ray Space
  Telescope (Fermi)} mission~\cite{Atwood:2009ez} to constrain the DM
density profile and the effect the uncertainties in the DM
distribution could have to constrain DM properties, as annihilation 
cross section, mass and branching ratio into dominant annihilation
channels.  We consider the potential gamma-ray signal in a squared
region with a side of $20^\circ$ around the GC ($|l|, |b| < 10^\circ$)
and, in order to model the relevant gamma-ray foregrounds, we use the
numerical code GALPROP~\cite{Strong:1998pw} and the latest {\it
  Fermi}--LAT observations~\cite{Collaboration:2010ru, Abdo:2010nz}.

During the last years, different approaches have been proposed to
determine the WIMP DM properties by using indirect or direct
measurements or their combination~\cite{Dodelson:2007gd, Bernal:2008zk,
  Bernal:2008cu, Jeltema:2008hf, Bernal:2010ip, Bernal:2010ti,
  PalomaresRuiz:2010pn, PalomaresRuiz:2010uu, Edsjo:1995zc,
  Cirelli:2005gh, Mena:2007ty, Agarwalla:2011yy,Das:2011yr,
  Lewin:1995rx, Primack:1988zm, Green:2007rb, Bertone:2007xj,
  Shan:2007vn, Drees:2008bv, Green:2008rd, Beltran:2008xg, Shan:2009ym,
  Strigari:2009zb, Peter:2009ak, Chou:2010qt, Shan:2010qv, Fox:2010bu,
  Kelso:2010sj, Bergstrom:2010gh, Pato:2010zk, Billard:2010jh,
  Shan:2011jz, Shan:2011ka}.  In addition, the information that could
be obtained from collider experiments would also be of fundamental
importance to learn about the nature of DM~\cite{Drees:2000he,
  Polesello:2004qy, Battaglia:2004mp, Allanach:2004xn,
  Weiglein:2004hn, Birkedal:2005jq, Moroi:2005zx, Nojiri:2005ph,
  Baltz:2006fm, Arnowitt:2007nt, Cho:2007qv, Arnowitt:2008bz,
  Belanger:2008yc, Cho:2008tj, Baer:2009bu, Feldman:2011me} and could
also be further constrained when combined with direct and indirect
detection data~\cite{Bernal:2008zk, Baltz:2006fm, Bourjaily:2005ax,
  Altunkaynak:2008ry, Bertone:2010rv}.  Therefore, in order to show
how well the DM density profile could be determined by the combination
of different type of experiments, we will also make use of potential
information that might be obtained on some parameters by other
complementary means in addition to the gamma-ray signal.

The paper is organized as follows.  In Section~\ref{gamma-rays} we
describe the most relevant (for this work) component of the gamma-ray
emission from DM annihilation in the GC.  In Section~\ref{MWprofile} we
state some of the results obtained in numerical N-body simulations
regarding the density profiles of CDM halos and we also briefly review
the current knowledge of some of the most important parameters that
determine the structure of the Milky Way halo.  We describe the
ingredients of our analysis in Section~\ref{analysis} and present the
results in Section~\ref{results}. Finally, we draw our conclusions in
Section~\ref{conclusions}.

\section{Gamma-rays from DM annihilations}
\label{gamma-rays}

The differential flux of prompt gamma--rays generated from DM
annihilations in the smooth Milky Way DM halo~\footnote{Throughout
  this work we neglect the contribution due to substructure in the
  Milky Way halo, which could enhance the gamma-ray flux from DM
  annihilation by a factor of $\lesssim$10~\cite{Diemand:2007qr,
    Diemand:2008in, Springel:2008cc}.} after the hadronization,
fragmentation and decay of the final states and coming from a
direction within a solid angle $\Delta\Omega$ can be written
as~\cite{Bergstrom:1997fj}
\begin{equation}
\left(\frac{d\Phi_{\gamma}}{dE_\gamma}\right)_{{\rm prompt}} (E_{\gamma},
\Delta\Omega) =  R_\odot\,\rho_\odot^2 \,\overline{J}(\Omega) \,
\frac{\Delta\Omega}{4\,\pi} \, \frac{\langle\sigma
  v\rangle}{2\,m_\chi^2} \, \sum_i \text{BR}_i \,
\frac{dN_{\gamma}^i}{dE_{\gamma}} ~,
\label{Eq:promptflux}
\end{equation}
where $R_\odot$ is the distance from the Sun to the GC, $\rho_\odot =
\rho(R_\odot)$ is the local DM density, $\langle\sigma v\rangle$ is
the thermal average of the total annihilation cross section times the
relative velocity, $m_\chi$ is the DM mass, the discrete sum is over
all DM annihilation channels, BR$_i$ is the branching ratio of DM
annihilation into the $i$-th final state and
$dN_{\gamma}^i/dE_{\gamma}$ is the differential gamma--ray yield of SM
particles into photons for the $i$-th channel, that we simulate with
the event generator PYTHIA 6.4~\cite{Sjostrand:2006za}, which
automatically includes the so-called final state radiation (photons
radiated off the external legs).  The dimensionless quantity
$\overline{J}$, which depends crucially on the DM distribution, is
defined as
\begin{equation}
\overline{J}(\Omega) = \frac{1}{\Delta\Omega} \,
\frac{1}{R_\odot\,\rho_\odot^2} \, \int_{\Delta\Omega}d\Omega \,
\int_\text{los} \rho\big(r(s,\Omega)\big)^2\ ds \, ~, 
\label{Eq:Jbarr} 
\end{equation}
where the spatial integration of the square of the DM density profile
$\rho(r)$ is performed along the line of sight within the solid angle of
observation $\Delta\Omega$.  More precisely, the distance from the GC is
$r = \sqrt{R^2_\odot -  2\,s\,R_\odot \cos \psi + s^2}$, and the upper
limit of integration is $s_{\rm max} = \sqrt{(R_{\rm vir}^2 - \sin^2
  \psi\,R^2_\odot)} + R_\odot \cos \psi$, where $\psi$ is the angle
between the direction of the GC and that of observation.  The virial
radius of the Milky Way DM halo is defined as $R_{\rm vir} = \left[3\,
M_{\rm vir} / (4 \pi\,\Delta_{\rm vir}\,\rho_{\rm crit})\right]^{1/3}$,
where $M_{\rm vir}$ is the virial mass of the DM halo, $\rho_{\rm
  crit}$ is the critical density of the Universe and $\Delta_{\rm vir}
= 18 \pi^2 + 82\,(\Omega_{\rm m}-1) - 39\,(\Omega_{\rm m}-1)^2$ is the
virial overdensity~\cite{Bryan:1997dn} with $\Omega_{\rm m} =
0.258$~\cite{Komatsu:2008hk} the non-relativistic matter contribution
to the total density of the Universe.  It is also customary to define
the concentration parameter as $c_{\rm vir} \equiv R_{\rm vir}/r_s$,
where $r_s$ is the scale radius.

\section{Milky Way DM density profile}
\label{MWprofile}

During the past two decades, the improvement in numerical simulations
has led to the discovery of a universal internal structure for
spherically averaged DM halos~\cite{Navarro:1995iw, Navarro:1996gj}
that can be well described by a density profile (NFW parametrization)
with two (free) parameters given by 
\begin{equation}
\rho(r) = \rho_\odot \, \frac{\left(R_\odot/r_s\right)
  \left[1+(R_\odot/r_s)\right]^2}{(r/r_s)\,[1+(r/r_s)]^2} ~.
\label{Eq:NFW}
\end{equation}
However, further simulations with
improved numerical resolution started to show some systematic
differences, mainly in the central regions of CDM halos, that have
created a vivid debate~\cite{Burkert:1995yz, Moore:1999gc,
  Ghigna:1999sn, Fukushige:1996nr, Subramanian:1999bg,
  Fukushige:2000ar, Klypin:2000hk, Taylor:2001bq, Power:2002sw, 
  Ricotti:2002qu, Fukushige:2003xc, Diemand:2004wh, Gentile:2004tb,
  Salucci:2007tm}.  The recent Via Lactea II
simulations~\cite{Diemand:2008in} seem to partly verify earlier
results, with a NFW density profile.  On the other hand, the Aquarius
project simulations~\cite{Springel:2008cc} seem to favor a slightly
different parametrization~\cite{Navarro:2003ew, Graham:2005xx,
  Graham:2006ae, Graham:2006af, Navarro:2008kc}, which is less cuspy
towards the center of the Galaxy than the NFW profile and has three
(free) parameters, the so-called Einasto profile~\cite{Einasto}, 
\begin{equation}
 \rho(r) = \rho_\odot \, \exp\left[-\frac{2}{\alpha}
   \left(\left(\frac{r}{r_s} \right)^\alpha -
   \left(\frac{R_\odot}{r_s} \right)^\alpha \right)\right] ~,  
\label{Eq:einasto}
\end{equation}
which is the same relation between slope and radius that defines
Sérsic's model~\cite{Sersic68}, but applied to space density instead
of the projected surface density of galaxies~\cite{Merritt:2005xc}.

In order to observationally determine the parameters of the DM density
distribution, different constraints are considered. For instance,
several dynamical tracers have been used to determine the Milky Way
mass~\footnote{More precisely, these analyses estimate the mass up to
  some radius, typically tens of kpc, and then infer the total mass
  from a model of the Milky Way.}, and most estimates have come from
escape velocity arguments or from Jeans modeling of the radial
density and velocity dispersion profiles of distant halo tracer
populations, as satellite galaxies, globular clusters and horizontal
branch halo stars on far orbits.  However, while the mass of external
galaxies can be determined with reasonable precision, the mass of the
Milky Way remains uncertain within a factor of $\sim$~6 ($5 \times
10^{11} \, {\rm M}_\odot \lesssim M_{\rm MW} \lesssim 3 \times 10^{12}
\, {\rm M}_\odot$)~\cite{Kochanek:1995xv, Wilkinson:1999hf,
  Sakamoto:2002zr, Smith:2006ym, Battaglia:2005rj, Dehnen:2006cm,
  Li:2007eg, Xue:2008se, Catena:2009mf, Weber:2009pt, Watkins:2010fe,
  Gnedin:2010fv, Przybilla:2010gd, McMillan:2011wd} due to the lack of
observational data for luminous tracers in its outer regions where DM
dominates.  Within the innermost $\sim 50-60$~kpc of the Milky Way, on
the high mass end we find $M_{\rm MW} (<50~{\rm kpc}) \equiv M_{\rm
  MW} (50) = \left(5.4 ^{+0.2}_{-3.6}\right) \times 10^{11} \, {\rm
  M}_\odot$~\cite{Wilkinson:1999hf}, whereas a lower mass was recently
obtained by estimating the circular velocity curve up to 60~kpc,
$M_{\rm dh} (<60~{\rm kpc}) \equiv M_{\rm dh} (60) = (4.0 \pm 0.7)
\times 10^{11} \, {\rm M}_\odot$~\cite{Xue:2008se}.  Here $M_{\rm MW}
(r)$ and $M_{\rm dh} (r)$ refer to the total mass (dark and visible)
of the Milky Way and the DM halo mass within a radius $r$,
respectively.  Below we will compare these estimates with the masses for
the profiles we consider, but we stress that the former estimate
refers to the total mass within a given radius and not only to the
mass of the DM halo.  The galactic disc and bulge (visible matter) are
estimated to contribute with a total mass about an order of magnitude
lower, $\sim (5-7) \times 10^{10} M_\odot$~\cite{Gerhard:2002ud,
  Flynn:2006tm, McMillan:2011wd}, which would bring the former
estimate closer to the latter one. 
 
The local DM density ($\rho_\odot$) is another important parameter in
the description of the DM halo.  This is, as well, a crucial parameter
in direct DM detection experiments and in the search for neutrinos
produced by DM annihilations at the center of the Sun or the Earth.
Commonly, the way to estimate the local DM density follows the
approach of Ref.~\cite{Caldwell:1981rj} and it is based in constructing
three-component (disc, bulge and dark halo) models for the galaxy and 
confront them against observational data.  It is usually assumed to be
$\rho_\odot = 0.3$~GeV/cm$^3$, although approximately with a factor of
two of uncertainty.  However, this estimate is just a rough central
value within the range obtained by several analyses from old dynamical
constraints and using cored DM profiles~\cite{Gates:1995dw}.
Moreover, it is curious to note that in these early estimates, to the
best of our knowledge, there is no justification for using that
particular value, which is not even an unweighted average, but on the
contrary, $\rho_\odot$ was usually quoted to be slightly larger, e.g.,
$\rho_\odot = 0.51 ^{+0.21}_{-0.17}$~GeV/cm$^3$~\cite{Gates:1995dw}.

Recently, a few new analyses have been performed using new data.  Some
of them follow the approach of Ref.~\cite{Caldwell:1981rj} but with a
large set of new observational constraints of the Milky
Way~\cite{Catena:2009mf, Weber:2009pt, Strigari:2009zb, deBoer:2010eh,
McMillan:2011wd} and in Ref.~\cite{Salucci:2010qr} a new technique is
proposed such that only local observables are used and there is no
need of a model for the galaxy.  In general, they tend to obtain
slightly larger values than the canonical one, $\rho_\odot = 0.39 \pm
0.03$~GeV/cm$^3$~\cite{Catena:2009mf}, $\rho_\odot = 0.32 \pm 
0.07$~GeV/cm$^3$~\cite{Strigari:2009zb}, $\rho_\odot = 0.43 \pm
0.15$~GeV/cm$^3$~\cite{Salucci:2010qr}, $\rho_\odot = 0.40 \pm
0.04$~GeV/cm$^3$~\cite{McMillan:2011wd}.  Let us also note that DM 
halos are not predicted to be perfectly spherical and this possible 
flattening could imply a larger DM local density~\cite{Gates:1995dw,
  Pato:2010yq}.  In this work we will take $\rho_\odot =
0.4$~GeV/cm$^3$ as our default value.

Correlated with the use of the DM local density is the knowledge of
the distance to the GC ($R_\odot$), which is also an important
parameter to determine the structure of the Milky Way.  It is usually 
assumed that $R_\odot = 8.5$~kpc.  However, this is based on an old
recommendation by the International Astronomical Union, after the
value for the unweighted average of $R_\odot$ obtained using different
estimates between 1974 and 1986~\cite{Kerr:1986hz}.  Nevertheless,
after a more careful statistical analysis of these results, which
tried to account for statistical and systematic errors and the
covariance among the different methods, a lower value of $R_\odot =
8.0 \pm 0.5$~kpc was proposed almost twenty years
ago~\cite{Reid:1993fx}.  Many new measurements were performed during
the following decade, with an average value $R_\odot = 7.80 \pm
0.33$~kpc~\cite{Avedisova:2005}, consistent within errors with
previous estimates.

During the last years, and using new astrometric and radial velocity
measurements of stellar orbits around the massive black hole in the
GC, new analyses seem to point to a slightly larger value, $R_\odot =
8.4 \pm 0.4$~kpc~\cite{Ghez:2008ms}, $R_\odot = 8.33 \pm
0.35$~kpc~\cite{Gillessen:2008qv},  $R_\odot = 8.2 \pm
0.5$~kpc~\cite{Bovy:2009dr}.  In addition, based on trigonometric
parallax observations of star forming regions, a similar result has
been derived, $R_\odot = 8.4 \pm 0.6$~kpc~\cite{Reid:2009nj} (see
however Ref.~\cite{McMillan:2009yr}).  All in all, even with these
seemingly converging results, recent measurements still span in the
range between $R_\odot = 7.2 \pm 0.3$~kpc~\cite{Bica:2005en} and
$R_\odot = 8.7 ^{+0.57}_{-0.43}$~kpc~\cite{Vanhollebeke:2009ka}. In
this work, we will take $R_\odot = 8.3$~kpc as our default
value. \\[2ex] 

From Eqs.~\eqref{Eq:NFW} and~\eqref{Eq:einasto}, we see that the DM
density profile is determined by two parameters in the NFW case
($\rho_\odot, \, r_s$) and three for the Einasto profile ($\rho_\odot,
\, r_s, \, \alpha$).  Nevertheless, one could equivalently use any set
of (free) parameters to describe the DM density profile with a given
parametrization.  For instance, in N-body simulations, it is usually
convenient to use the virial mass and the concentration parameter
instead of the local DM density and the scale radius.  On the other
hand, it turns out that different analyses, which make use of the
results from N-body simulations, have shown that the structural
properties of DM halos depend on the halo mass~\cite{Navarro:1996gj,
  Jing:1999km, Bullock:1999he, Wechsler:2001cs, Maccio':2006nu,
  Maccio':2008xb, Eke:2000av, Zhao:2002rm, Zhao:2003jf, Neto:2007vq,
  Gao:2007gh, Duffy:2008pz, Zhao:2008wd, Klypin:2010qw,
  MunozCuartas:2010ig}.  In general, the correlations among different
parameters also embed the dependence on the cosmological parameters.
In particular, these analyses have found an anti-correlation between
$M_{\rm vir}$ and $c_{\rm vir}$, which is especially sensitive to the
value of $\sigma_8$ (a measurement of the amplitude of the linear power
spectrum on the scale of 8~$h^{-1}$~Mpc).  The median relation between
$M_{\rm vir}$ and $c_{\rm vir}$ (assuming a NFW profile) at redshift
$z=0$ and for relaxed halos~\footnote{These are halos in dynamical
  equilibrium, in general with smooth density profiles and reasonably
  well fitted by NFW (and Einasto) profiles.  However, the degree of
  relaxedness could be very important to determine halo
  properties~\cite{Skibba:2011mj}.}, using the cosmological parameters  
from the 5-year release of the WMAP mission~\cite{Komatsu:2008hk}, is
given by~\cite{MunozCuartas:2010ig}
\begin{equation}
c_{\rm vir} = 10^{2.155} \, \left( \frac{M_{\rm vir}}{h^{-1} {\rm
    M}_\odot} \right)^{-0.097} = 10.11 \, \left( \frac{M_{\rm
    vir}}{10^{12} \, {\rm M}_\odot} \right)^{-0.097}~, 
\label{Eq:cM}
\end{equation}
where $h=0.72$~\cite{Komatsu:2008hk}.  Hence, this type of correlation
would reduce further the number of parameters to describe the DM
density profile to one (to two in the analogous case of the Einasto
profile).  However, it is important to bare in mind that
Eq.~\eqref{Eq:cM} represents the mean concentration for a given virial
mass, so it does not account for the intrinsic scatter about these mean
relation~\cite{Jing:1999km, Bullock:1999he, Wechsler:2001cs,
  Maccio':2006nu, Maccio':2008xb}.  The distribution of the logarithm
is approximately Gaussian and at $1\sigma$ is typically $\sigma_{\ln
  c_{\rm vir}} \simeq 0.3$~\cite{Jing:1999km, Bullock:1999he,
  Wechsler:2001cs, AvilaReese:2005fm, Maccio':2006nu, Maccio':2008xb}
(see however Ref.~\cite{Reed:2010gh} where instead of a log-normal
distribution of halo concentrations, a Gaussian distribution is found
and where also the uncertainty in DM annihilation from the Milky Way
halo that arises from the distribution of halo concentrations is
discussed).  In the analysis that follows, we will also consider the
possibility that the Milky Way satisfies this mean relation
(Eq.~\eqref{Eq:cM}) obtained in numerical N-body simulations.

\section{Analysis}
\label{analysis}

The {\it Fermi} telescope~\cite{Atwood:2009ez} was launched in June
$2008$ for a mission of 5~to 10~years.  {\it Fermi}--LAT is the
primary instrument on board of the {\it Fermi} mission and it performs
an all-sky survey with a field of view $\text{FoV}=2.4$~sr, covering
an energy range from below 20~MeV to more than $300$~GeV for
gamma-rays, with an energy and angle-dependent effective area which is
approximately $A_\text{eff} \simeq 8000$~cm$^2$.  We consider a
constant effective area, but we notice that different effects could
reduce the exposure after a more detailed description of the detector
is considered, which however is beyond the scope of this paper.  {\it
  Fermi}--LAT's  equivalent Gaussian $1\sigma$ energy resolution is
$\sim 10\%$ at energies above 1~GeV.  On the other hand, for normal
incidence, 68\% of the photons at 10~GeV are contained within $\sim
0.25^\circ$.  In the following analysis, we consider a 5-year mission
run, an energy interval from 1~GeV to 300~GeV and a square region of
observation with a side of $20^\circ$ around the GC ($|l|, |b| <
10^\circ$).  In our simplified modeling of the detector, we do not
explicitly compute the effects of the energy resolution and of the
point spread function.  However, in order to avoid that this affects
significantly our results, we divide the energy interval into 20
evenly spaced logarithmic bins and consider 10 concentric $1^\circ$
bins.  Our aim is to illustrate the issues here discussed and serve as
a starting analysis for more detailed ones.

In order to study the capabilities of the {\it Fermi}--LAT experiment
to simultaneously constrain some DM particle properties and the Milky
Way DM density profile, we begin by modeling the background. We
consider the three components contributing to the high-energy
gamma-ray background: the diffuse galactic emission (DGE), the
contribution from resolved point sources (PS) and the isotropic
gamma-ray background (IGRB).  We model the DGE with the GALPROP code 
(v54)~\cite{Strong:1998pw} by means of the WebRun
service~\cite{Vladimirov:2010aq} and we take the conventional   
model, which reproduces reasonably well the recent measurements by
{\it Fermi}--LAT, at least at intermediate galactic latitudes
$10^\circ < |b| < 20^\circ$ and up to 10~GeV~\cite{Abdo:2009ka,
  Abdo:2009mr}.  The other two sources of background are taken from
the {\it Fermi}--LAT published results~\cite{Collaboration:2010ru,
  Abdo:2010nz}.  We refer the reader to Ref.~\cite{Bernal:2010ip} for
further details on the gamma--ray foregrounds we consider.  However,
let us note that Ref.~\cite{Bernal:2010ip} makes use of the {\it
  Fermi} Science Tools, both to model the background and the signal.
We have checked that our simplified modeling of the detector and
backgrounds does not introduce significant changes in our results and
allow us to substantially reduce the time of computation without
modifying our conclusions.  

During the last years, different approaches have been considered to
constrain WIMP DM properties by using indirect searches, direct
detection measurements, collider information or their
combination~\cite{Dodelson:2007gd, Bernal:2008zk, Bernal:2008cu, 
  Jeltema:2008hf, Bernal:2010ip, Bernal:2010ti, PalomaresRuiz:2010pn,
  PalomaresRuiz:2010uu, Edsjo:1995zc, Cirelli:2005gh, Mena:2007ty,
  Agarwalla:2011yy, Das:2011yr, Lewin:1995rx, Primack:1988zm,
  Green:2007rb, Bertone:2007xj, Shan:2007vn, Drees:2008bv,
  Green:2008rd, Beltran:2008xg, Shan:2009ym, Strigari:2009zb,
  Peter:2009ak, Chou:2010qt, Shan:2010qv, Fox:2010bu, Kelso:2010sj,
  Bergstrom:2010gh, Pato:2010zk, Billard:2010jh, Shan:2011jz,
  Shan:2011ka, Drees:2000he, Polesello:2004qy, Battaglia:2004mp,
  Allanach:2004xn, Weiglein:2004hn, Birkedal:2005jq, Moroi:2005zx,
  Nojiri:2005ph, Baltz:2006fm, Arnowitt:2007nt, Cho:2007qv,
  Arnowitt:2008bz, Belanger:2008yc, Cho:2008tj, Baer:2009bu,
  Feldman:2011me, Bourjaily:2005ax, Altunkaynak:2008ry,
  Bertone:2010rv}.  These measurements are complementary and 
constitute an important step toward identifying the particle nature of 
DM.  However, astrophysical searches are hindered by large
uncertainties in the structure of the Milky Way DM halo and, although
different profiles have been considered, it is customary to assume a
given DM density profile in order to study the capabilities of
gamma-ray experiments to extract some DM
properties~\cite{Dodelson:2007gd, Bernal:2008zk, Bernal:2008cu,
  Jeltema:2008hf, Bernal:2010ip, Bernal:2010ti} (for attempts to also
fit the DM density profile see Ref.~\cite{Dodelson:2007gd}, but
assuming the annihilation channel to be known and
Refs~\cite{Bi:2009am, Liu:2009sq}, but for a very large annihilation 
cross section).

In this work, we discuss the {\it Fermi}--LAT's abilities to constrain
the DM density profile and some DM properties after 5~years of data
taking.  We evaluate the effect the uncertainties in the DM density
profile could have on the determination of the DM mass, annihilation
cross section and the annihilation channels.  Concerning the DM
density profile, when we consider an Einasto profile, we fit its three
parameters: the DM local density $\rho_\odot$, the scale radius $r_s$
and the index $\alpha$; whereas for the case of a NFW profile we fit
$c_{\rm vir}$ and $M_{\rm vir}$.  In principle, as with respect to DM
properties, the analysis should include all possible annihilation
channels, but this would render the present analysis very much time
consuming.  Nevertheless, in practice, they are commonly classified as
hadronic (soft channels) and leptonic channels (hard channels).  Thus,
for simplicity, when simulating a signal, we only consider two
possible (generic) channels, $\tau^+\tau^-$ or $b\bar
b$~\footnote{Note that for these channels, and for the energy range of
  interest here, the contribution from the electrons and positrons
  produced in DM annihilations to the gamma-ray spectrum via inverse
  Compton scattering off the ambient photon background, is subdominant
  with respect to prompt gamma-rays~\cite{Bernal:2010ip}.  Also note
  that for the considered masses, electroweak radiative corrections to
  DM annihilations with gauge boson bremsstrahlung do not
  contribute~\cite{Chen:1998dp, Kachelriess:2007aj, Bell:2008ey, 
    Dent:2008qy, Ciafaloni:2010qr, Kachelriess:2009zy, Yaguna:2010hn, 
    Ciafaloni:2010ti, Bell:2010ei, Bell:2011eu}.}.  By using 
this simplification, we can reduce the number of total free parameters
to six for the Einasto profile and five for the NFW profile: the mass,
$m_\chi$, the annihilation cross section, $\langle\sigma v\rangle$,
the branching ratio into channel 1, $\text{BR}_{1(2)}$ (or
equivalently into channel 2, $\text{BR}_{2(1)}=1-\text{BR}_{1(2)}$)
and the parameters describing the DM density profile $(\rho_\odot,
r_s, \alpha)$ or $(M_{\rm vir}, c_{\rm vir})$ for the Einasto or NFW
profile, respectively.  In order to focus on the determination of
these parameters and to avoid adding more parameters to the analysis,
we assume perfect knowledge of the background.  Hence, in this regard,
the results here presented should be taken as the most optimistic ones.
We use the $\chi^2$ function defined as
\begin{equation}
\chi^2  \left({\bf \theta}\right) = \sum_{j=1}^{10} \sum_{i=1}^{20}
\frac{\left( 
S_{ij}\left({\bf \theta}\right)  - 
S_{ij}^\text{th}\left({\bf \theta^0}\right) \right)^2}
{(\sigma_{\rm stat})_{ij}^2 + (\sigma_{\rm sys})_{ij}^2} ~, 
\label{Eq:chi2}
\end{equation}
where $S_{ij}({\bf \theta})$ represents the simulated signal events in
the $i$-th energy and $j$-th angular bin for each set of the
parameters ${\bf \theta} = \left(\rho_\odot, \, r_s, \, \alpha, \,
m_\chi, \, \langle\sigma v\rangle, \, \text{BR}_{\tau(b)} \right)$ for
the case of the Einasto profile or ${\bf \theta} = \left(M_{\rm vir},
c_{\rm vir}, \, m_\chi, \, \langle\sigma v\rangle, \, \text{BR}_{\tau(b)}
\right)$ for the NFW profile, and $S_{ij}^\text{th} ({\bf \theta^0})$ is
the assumed observed signal events in that bin with parameters 
${\bf \theta^0} = \left(\rho_\odot^0, \, r_s^0, \, \alpha^0, \,
m_\chi^0, \, \langle\sigma v\rangle^0, \, \text{BR}_{\tau(b)}^0 \right)$
for the Einasto profile or ${\bf \theta^0} = \left(M_{\rm vir}^0, \,
c_{\rm vir}^0, \, m_\chi^0, \, \langle\sigma v\rangle^0, \,
\text{BR}_{\tau(b)}^0 \right)$ for the NFW profile.  The statistical
errors are given by
\begin{equation}
(\sigma_{\rm stat})_{ij}^2 = S_{ij}^\text{th} \left({\bf \theta^0}
  \right) +B_{ij} ~, 
\label{Eq:staterror}
\end{equation}
where $B_{ij}$ is the number of background events in the $i$-th energy
and $j$-th angular bin.  On the other hand, in the analysis of
Ref.~\cite{Abdo:2010nz}, the rms of the residual count fraction between
{\it Fermi}--LAT data and the DGE model for energies above 200~MeV is
larger than the expected statistical errors by a factor of $\sim$~2.5.
Hence, baring in mind that the number of events in each bin is in
general dominated by the background events, we conservatively take the
systematic errors to be three times the statistical ones,
i.e., $\sigma_{\rm sys} = 3 \, \sigma_{\rm stat}$, in order to account
for the uncertainties in the determination of the background.  We
expect however that this will improve in the near future.

\section{Results}
\label{results}

\begin{table}[t]
\begin{center}
\begin{tabular}{|c|c|c|c|c|c|c|c|c|c|}
\cline{2-10}
\multicolumn{1}{c|}{}
  & $m_\chi$ & $\rho_\odot$ & $r_s$ & $\alpha$ & $c_{\rm vir}$ &
$M_{\rm dh} (50)$ & $M_{\rm dh} (60)$ & $M_{\rm vir}$ & Priors \\  
\multicolumn{1}{c|}{} & [GeV] & [GeV/cm$^3$] & [kpc] & & & $[10^{11}
  {\rm M}_\odot]$ & $[10^{11} {\rm M}_\odot]$ & $[10^{11} {\rm
    M}_\odot]$ & \\
\hline
Fig.~\ref{Fig:80} 
& 80 & 0.4 & 20 & 0.17 & 15.2 & 4.6 & 5.5 & 16 & No \\[1ex] 
Fig.~\ref{Fig:25}  
& 25 & 0.4 & 20 & 0.17 & 15.2 & 4.6 & 5.5 & 16 & No \\[1ex] 
Fig.~\ref{Fig:80var} 
& 80 & 0.4 & 20 & 0.17 & 15.2 & 4.6 & 5.5 & 16 & Yes \\[1ex] 
Fig.~\ref{Fig:25var}
& 25 & 0.4 & 20 & 0.17 & 15.2 & 4.6 & 5.5 & 16 & Yes \\[1ex]  
Fig.~\ref{Fig:25cnfw}
& 25 & 0.3 & 38.6 & - & 9.2 & 4.8  & 6.0 & 25 & No/Yes \\[1ex] 
Fig.~\ref{Fig:25bnfw} 
& 25 & 0.4 & 20 & - & 15.2 & 4.7 & 5.6 & 16 & No/Yes \\[1ex] 
Fig.~\ref{Fig:25anfw}  
& 25 & 0.4 & 10 & - & 24.4 & 3.5 & 4.0 & 8.4 & No/Yes \\
\hline
\end{tabular}
\caption{\sl \textbf{\textit{Summary of the parameters and relevant
      information for each of the figures in Section~\ref{results}}}.
  We take as default values: $R_\odot = 8.3$~kpc, DM annihilation into
  a pure $b \bar b$ final state and $\langle \sigma v \rangle = 3
  \times 10^{-26}$~cm$^3$/s.  The Einasto (NFW) profile is used when a
  value for $\alpha$ is (not) indicated.  When priors are added,
  $\sigma_{\log \langle \sigma v \rangle} = 0.15$ is always assumed
  and for $\rho_\odot$ two cases are considered: $\sigma_{\rho_\odot}=
  0.04$~GeV/cm$^3$ and $\sigma_{\rho_\odot}= 0.2$~GeV/cm$^3$.  All the
  figures assume 5 years of data taking.}
\label{Tab:figs}
\end{center}
\end{table}

In this section, we first consider the Einasto profile and show the
results for two different DM masses, $m_\chi = 25$~GeV and $m_\chi =
80$~GeV.  We show the capabilities of the {\it Fermi}--LAT telescope
to constrain the three parameters that describe the DM density
distribution.  In addition, we also show the effect our ignorance
about the halo profile has on the determination of some DM properties.
This could be compared with the results obtained in
Ref.~\cite{Bernal:2010ip}, where a known DM density profile was
assumed.  Next, we also study a NFW profile for $m_\chi = 25$~GeV for
different values of the parameters of the DM profile and plot the
results in the $(M_{\rm vir}, \, c_{\rm vir})$  plane.  In this way,
one can readily compare our results to the relations obtained in
numerical N-body simulations, e.g., to Eq.~\eqref{Eq:cM}.  Moreover, we
also study the impact of adding some priors on some of the relevant
parameters, which could be obtained in the future using different
data, as $\langle \sigma v \rangle$ and $\rho_\odot$.  In
Table~\ref{Tab:figs} we summarize the parameters used and the relevant 
information in each of the figures we describe below.

\subsection{Constraining DM properties and the DM density profile}

\begin{figure}[t]
\begin{center}
\includegraphics[width=8.1cm]{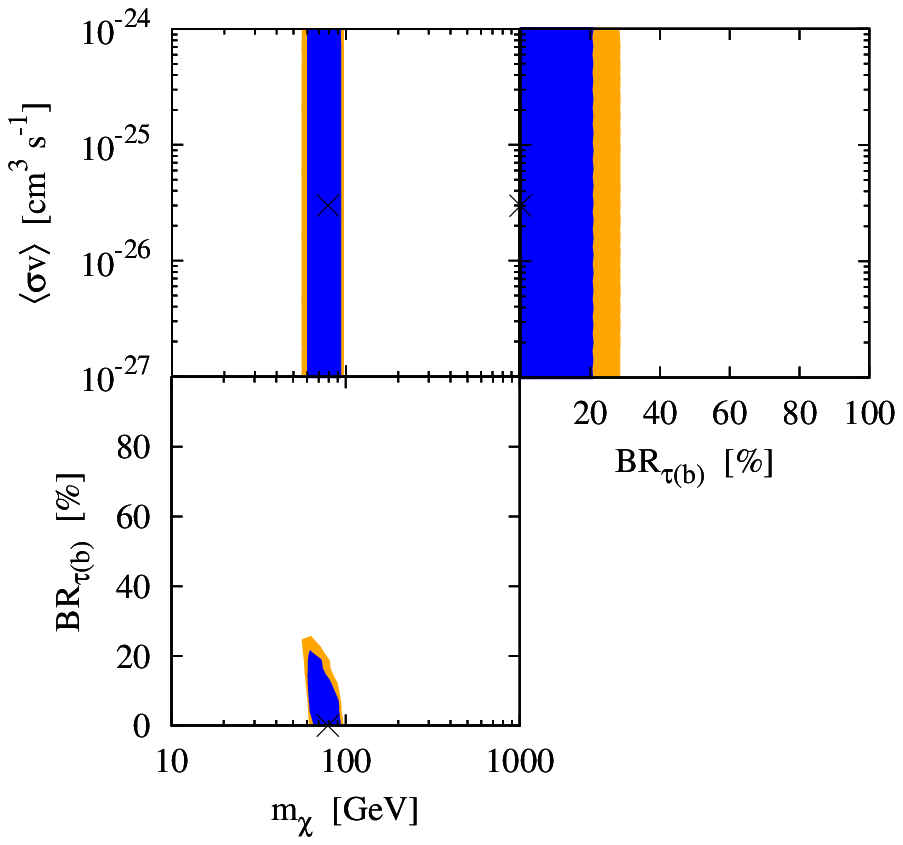}
\includegraphics[width=8.1cm]{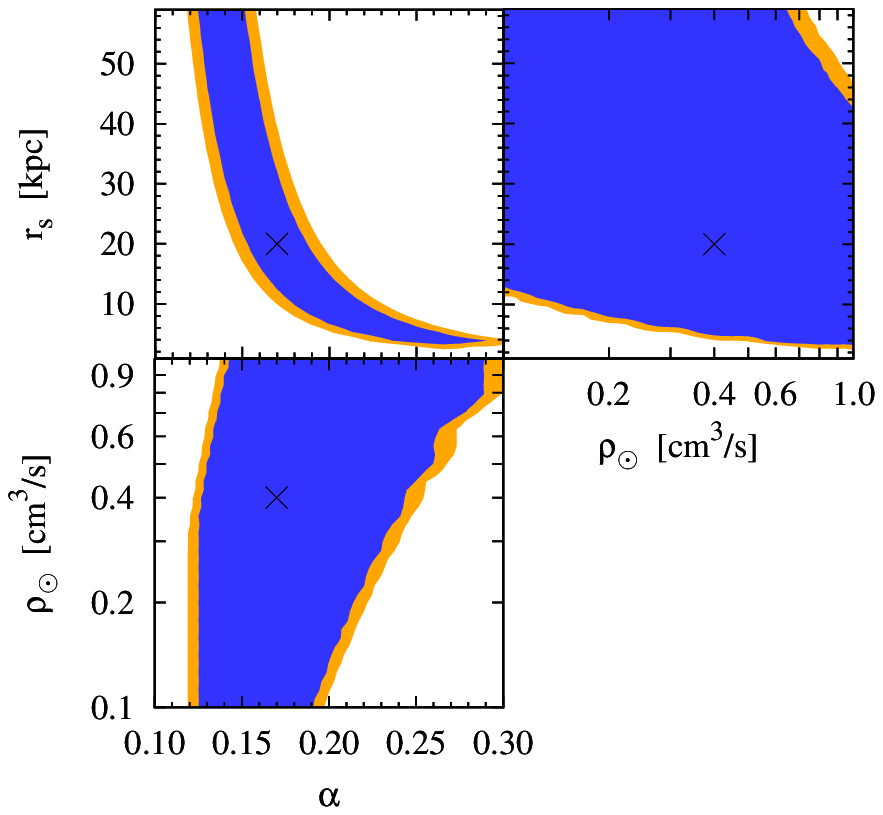}
\end{center}
\vspace{-0.8cm}
\caption{\sl \textbf{\textit{Fermi--LAT abilities to simultaneously
      constrain DM properties}} (left panel) \textbf{\textit{and the
      Milky Way DM density distribution}} (right panel)
    \textbf{\textit{after 5 years}}.  We consider DM annihilation into a
  pure $b \bar b$ final state ($\text{BR}_{\tau(b)} = 0$), $\langle
  \sigma v \rangle = 3 \times 10^{-26}$~cm$^3$/s and $m_\chi =
  80$~GeV.  For the DM density profile we consider an Einasto profile 
  with $\rho_\odot = 0.4$~GeV/cm$^3$, $r_s = 20$~kpc, $\alpha = 0.17$ 
  and $R_\odot = 8.3$~kpc. The black crosses indicate the values of
  these parameters. In both panels, blue (orange) regions represent
  the 68\% CL (90\% CL) contours for 2~dof.}
\label{Fig:80}
\end{figure}

\begin{figure}[t]
\begin{center}
\includegraphics[width=8.1cm]{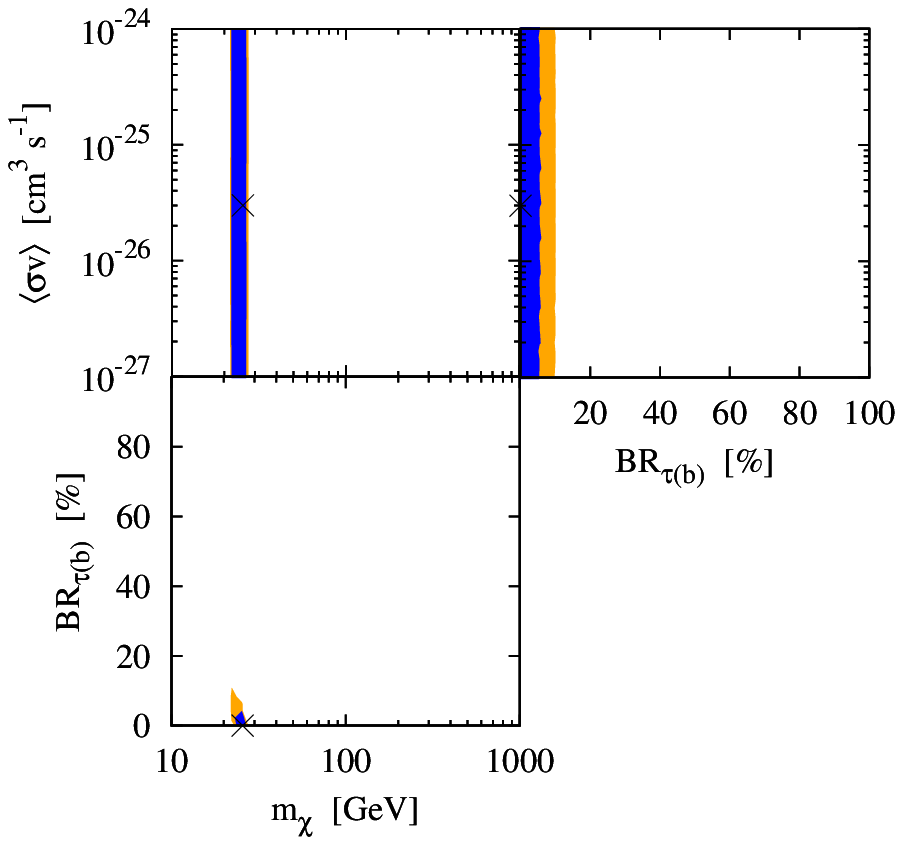}
\includegraphics[width=8.1cm]{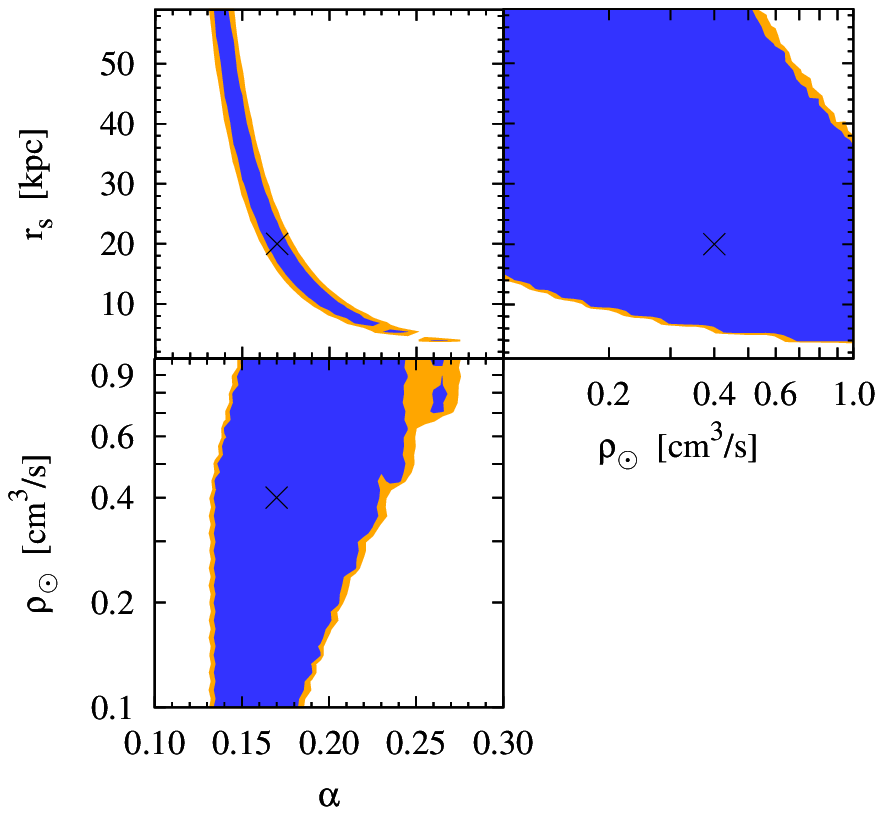}
\end{center}
\vspace{-0.8cm}
\caption{\sl \textbf{\textit{Fermi--LAT abilities to simultaneously
      constrain DM properties}} (left panel) \textbf{\textit{and the
      Milky Way DM density distribution}} (right panel)
    \textbf{\textit{after 5 years}}.  Same as Fig.~\ref{Fig:80} but
    for $m_\chi = 25$~GeV.}
\label{Fig:25}
\end{figure}

\begin{figure}[t]
\begin{center}
\includegraphics[width=8.1cm]{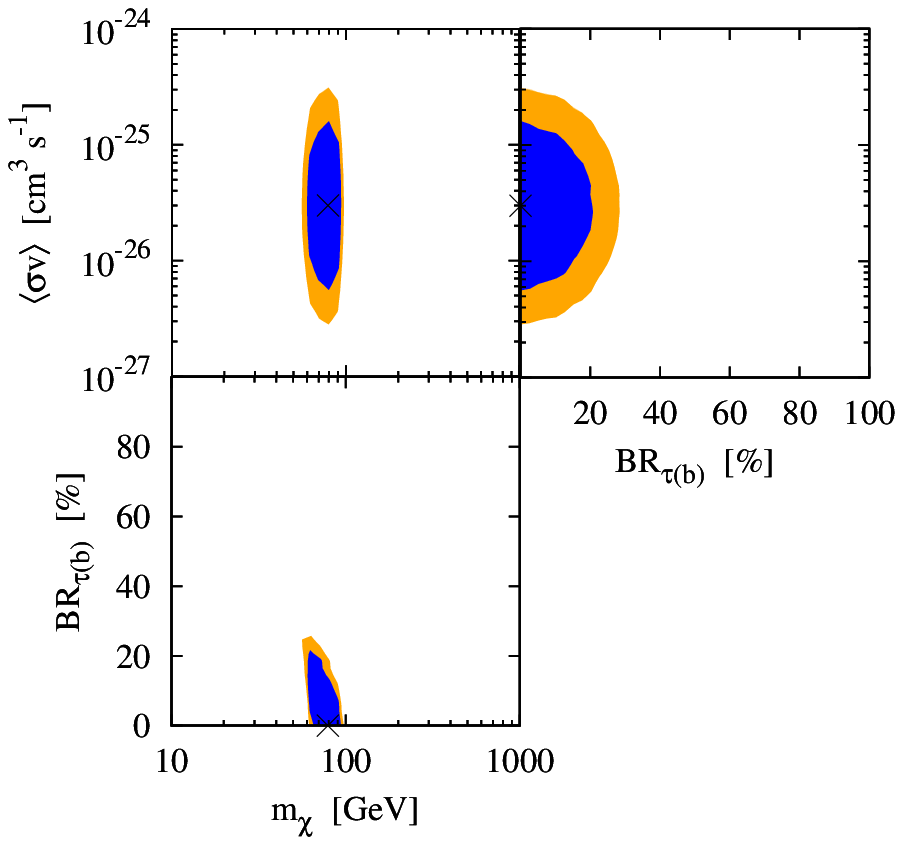}
\includegraphics[width=8.1cm]{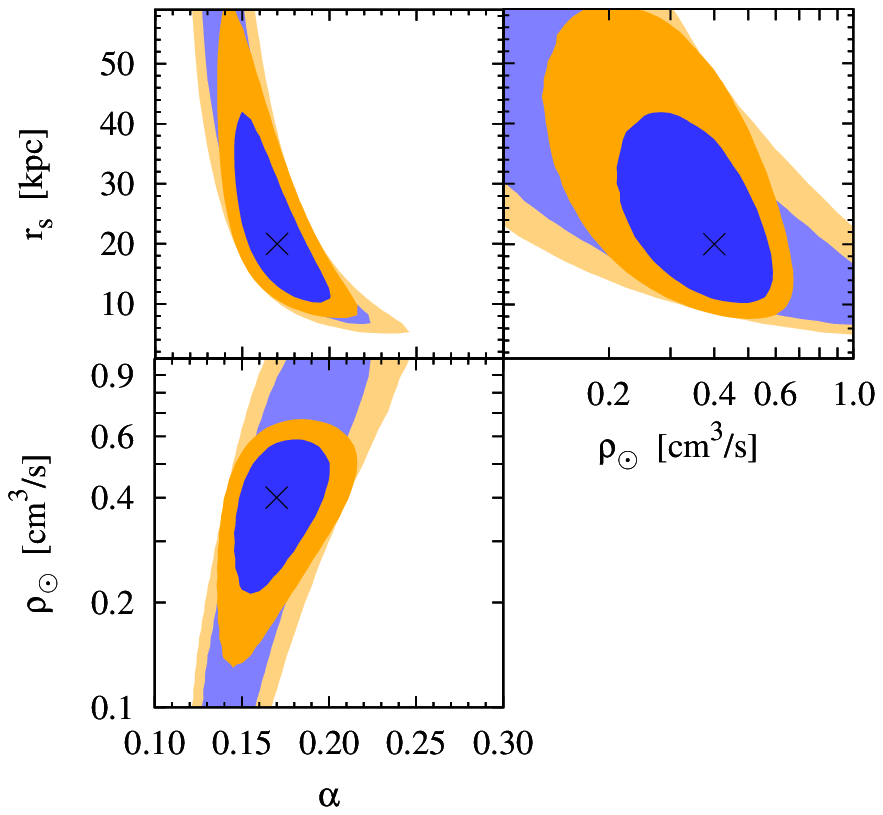}
\end{center}
\vspace{-0.8cm}
\caption{\sl \textbf{\textit{Fermi--LAT abilities to simultaneously
      constrain DM properties}} (left panel) \textbf{\textit{and the
      Milky Way DM density distribution}} (right panel)
  \textbf{\textit{after 5 years, with the addition of external
      information on $\boldsymbol{\rho_\odot}$ and
      $\boldsymbol{\langle \sigma v \rangle}$}} (see text).  Same
  parameters as in Fig.~\ref{Fig:80}.  In both panels, the dark
  (light) regions represent the case of $\sigma_{\log \langle \sigma
    v \rangle} = 0.15$ and $\sigma_{\rho_\odot}= 0.1 \, \rho_\odot$
  ($\sigma_{\rho_\odot}= 0.5 \, \rho_\odot$). As in Fig.~\ref{Fig:80},
  blue (orange) regions represent the 68\% CL (90\% CL) contours for
  2~dof.}
\label{Fig:80var}
\end{figure}

\begin{figure}[t]
\begin{center}
\includegraphics[width=8.1cm]{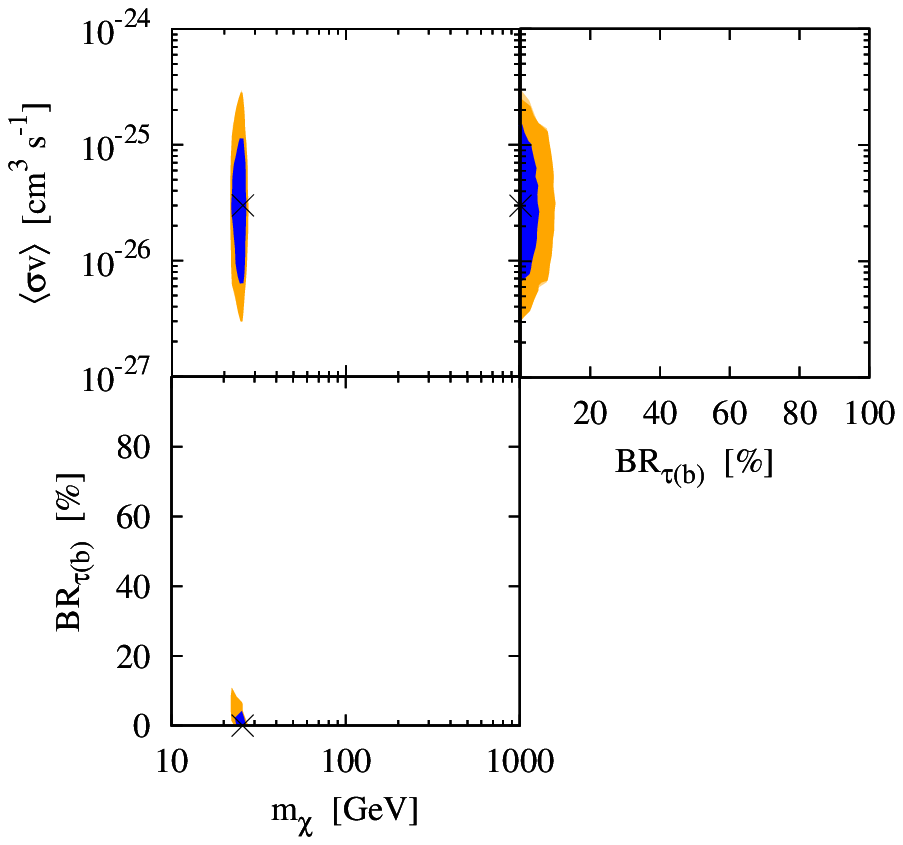}
\includegraphics[width=8.1cm]{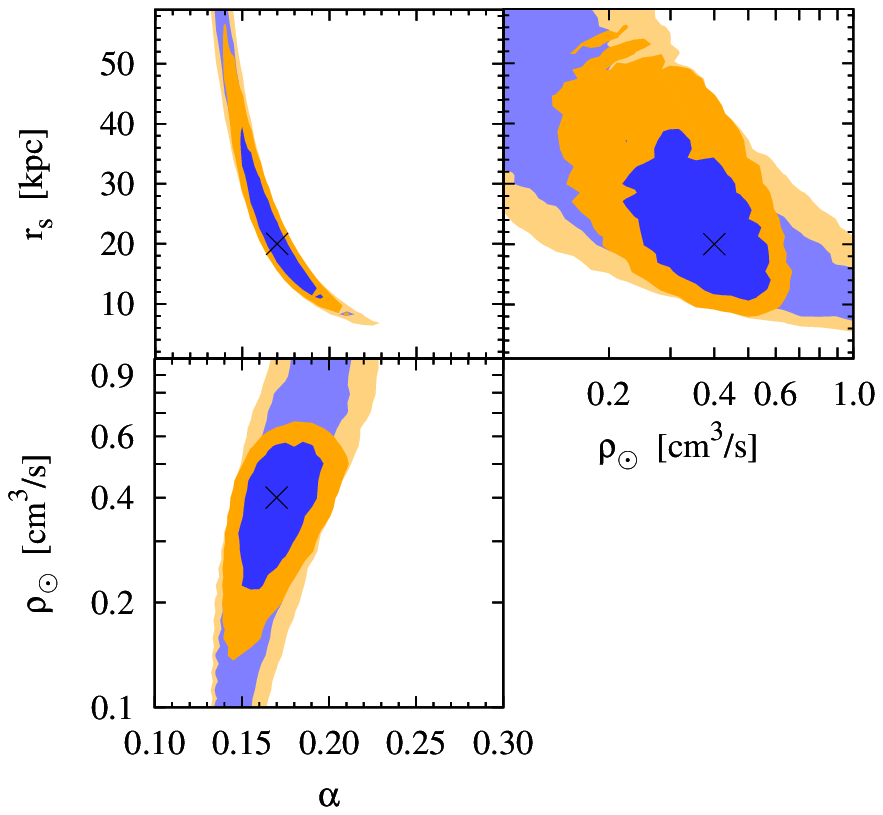}
\end{center}
\vspace{-0.8cm}
\caption{\sl \textbf{\textit{Fermi--LAT abilities to simultaneously
      constrain DM properties}} (left panel) \textbf{\textit{and the
      Milky Way DM density distribution}} (right panel)
  \textbf{\textit{after 5 years, with the addition of external
      information on $\boldsymbol{\rho_\odot}$ and
      $\boldsymbol{\langle \sigma v \rangle}$}} (see text).  Same as
  Fig.~\ref{Fig:80var} but for $m_\chi = 25$~GeV.}
\label{Fig:25var}
\end{figure}

For the first four figures, Figs.~\ref{Fig:80}--\ref{Fig:25var}, we
consider an Einasto profile (Eq.~\eqref{Eq:einasto}) with our default
values for the local DM density and the Sun distance to the GC
($\rho_\odot = 0.4$~GeV/cm$^3$ and $R_\odot = 8.3$~kpc) and we further
set the other two parameters to $r_s = 20$~kpc and $\alpha = 0.17$.
These values for the scale radius and the index $\alpha$ are the ones
commonly assumed in studies of DM indirect detection.  Moreover, this
value of $\alpha$ approximately represents the mean value obtained
from N-body simulations for dark halos with virial masses of the order
of $10^{12} \, {\rm M}_\odot$~\cite{Navarro:2008kc}.  Concerning the
DM particle properties, we assume DM annihilation into a pure $b \bar
b$ final state and an annihilation cross section (times relative
velocity) $\langle \sigma v \rangle = 3 \times 10^{-26}$~cm$^3$/s.
The left panels of the four figures show the results for pairs of DM
properties after marginalizing with respect to the rest of the
parameters and the right panels show the results for the parameters
that describe the DM density profile.

In Fig.~\ref{Fig:80}, we depict the {\it Fermi}--LAT reconstruction
prospects after 5 years.  The blue regions and the orange regions
correspond to the 68\% and 90\% confidence level (CL) contours for 2
degrees of freedom (dof), respectively.  However, note that there are
very small differences between these contours, which is due to a very
steep $\chi^2$.

From the left panel one can extract two important pieces of
information.  On one side, unlike what happens if the DM density 
profile is assumed to be known~\cite{Bernal:2010ip}, in the present
case the value of the annihilation cross section would remain
completely unknown.  This can be easily understood by the fact that by
also fitting the DM density profile, the quantity $\rho_\odot^2 \,
\overline{J}$ is allowed to vary and along with $\langle \sigma v
\rangle$, that quantity sets the normalization of the potential signal
(see~Eq.~\eqref{Eq:promptflux}).  Thus, there is a clear degeneracy
between them, which cannot be broken by the energy or angular binning
of the signal.  On the other hand, the capabilities for the
reconstruction of the DM mass and annihilation channel do not
substantially worsen with respect to the case of a known DM density 
profile (but with a single angular bin)~\cite{Bernal:2010ip}.  This
is clearly due to binning the signal both spatially and in energy, as
there are more ``data'' points and the energy spectrum obviously
depends on the DM mass and on the annihilation channel.  Hence, the DM
mass could be determined with a $\sim 25\%$ uncertainty and the
branching ratio could simultaneously be constrained to be
$\text{BR}_{\tau(b)} \lesssim 0.25$ at 90\% CL (2~dof). 

From the right panel of Fig.~\ref{Fig:80}, one realizes that it could
be possible to establish a strong correlation between $r_s$ and
$\alpha$ at a high CL (note again that the $\chi^2$ is very steep).
It is important to stress that this could be achieved by only using
gamma-ray observations with no further constraints on the relevant
parameters and basically thanks to the angular binning.  There are
also correlations among these parameters and $\rho_\odot$, although
they are much weaker.  Nevertheless, due to the degeneracy between
$\rho_\odot^2 \, \overline{J}$ and $\langle \sigma v \rangle$, it
turns out that the actual values of the parameters that describe the
DM density profile cannot be determined with only gamma-rays.
Although being a very weak constraint, $\alpha$ could be found to lie
in the range $\alpha = (0.12-0.30)$ at 90\% CL (2~dof).

In Fig.~\ref{Fig:25} we show the corresponding results for the case of
$m_\chi = 25$~GeV.  The same trends as in Fig.~\ref{Fig:80} are
observed, although due to larger statistics, slightly better results
would be obtained.  We see that the DM mass could be determined with a
$\sim 10\%$ uncertainty and simultaneously, the branching ratio could
be constrained to be $\text{BR}_{\tau(b)} \lesssim 0.10$ at 90\% CL
(2~dof).  In addition, the index $\alpha$ could be bounded to lie in
the range $\alpha \simeq (0.13-0.27)$ at 90\% CL (2~dof).  We also
note that the correlations between parameters get tighter, but the
degeneracy between $\rho_\odot^2 \, \overline{J}$ and $\langle \sigma
v \rangle$ cannot be broken, which would prevent us from being able to
determine neither the annihilation cross section nor the exact
parameters that describe the DM density profile.

So far, we have shown results that could be obtained by just using
gamma-ray observations.  However, we already have some information about
some of the parameters entering the fit, as the local DM density
($\rho_\odot$)~\cite{Catena:2009mf, Weber:2009pt, Strigari:2009zb, 
  deBoer:2010eh, McMillan:2011wd, Salucci:2010qr}, and one would also
hope that other parameters might be constrained in the future, as the
annihilation cross section ($\langle \sigma v \rangle$) and the DM
mass ($m_\chi$)~\cite{Lewin:1995rx, Primack:1988zm, Green:2007rb,
  Bertone:2007xj, Shan:2007vn, Drees:2008bv, Green:2008rd,
  Beltran:2008xg, Shan:2009ym, Strigari:2009zb, Peter:2009ak,
  Chou:2010qt, Shan:2010qv, Fox:2010bu, Kelso:2010sj,
  Bergstrom:2010gh, Pato:2010zk, Billard:2010jh, Shan:2011jz,
  Shan:2011ka, Drees:2000he, Polesello:2004qy, Battaglia:2004mp,
  Allanach:2004xn, Weiglein:2004hn, Birkedal:2005jq, Moroi:2005zx,
  Nojiri:2005ph, Baltz:2006fm, Arnowitt:2007nt, Cho:2007qv,
  Arnowitt:2008bz, Belanger:2008yc, Cho:2008tj, Baer:2009bu,
  Feldman:2011me, Bourjaily:2005ax, Altunkaynak:2008ry,
  Bertone:2010rv}.  In Figs.~\ref{Fig:80var}~and~\ref{Fig:25var}, we
show how our knowledge about the structure of the Milky Way dark halo
could be largely improved if we add to the gamma-ray analysis,
constraints coming from other sources on $\rho_\odot$ and $\langle
\sigma v \rangle$.

For the local DM density we will consider two different priors.  On
one side, we follow the common and conservative approach of assuming
that it is determined within a factor of two and set a 50\% error, 
$\sigma_{\rho_\odot}= 0.5 \, \rho_\odot$.  On the other hand, we also
consider a more optimistic constraint as recently obtained in different
works~\cite{Catena:2009mf, McMillan:2011wd}, $\sigma_{\rho_\odot}=
0.1 \, \rho_\odot$.

Concerning some of the DM properties, we will only set a prior on the
annihilation cross section.  Although there are good prospects of
constraining the DM mass by using direct searches and collider 
experiments, we will not implement this potential constraint in what
follows.  As shown above, a gamma-ray observation could allow a quite
good determination of the DM mass if the DM particle is relatively
light ($m_\chi \lesssim 100$~GeV).  Indeed, we have checked that a
further external constraint on this parameter would not add
substantial information.  However, as also discussed above, there is a 
degeneracy between $\rho_\odot^2 \, \overline{J}$ and $\langle \sigma
v \rangle$, so in order to break it as much as possible, not only is
the information on $\rho_\odot$ important, but a constraint on $\langle
\sigma v \rangle$ is also needed.  Hence, we assume that the
annihilation cross section might be determined within an order of
magnitude at $3\sigma$ with future collider data~\cite{Baltz:2006fm}
and set a very optimistic prior $\sigma_{\log \langle \sigma v
  \rangle} = 0.15$, where $\langle \sigma v \rangle$ is expressed in  
cm$^3$/s.  We note that a weaker constraint on this parameter would
not improve significantly the results as compared to the case with no
priors, so in this regard, information on both $\rho_\odot$ and 
$\langle \sigma v \rangle$ is crucial to reduce the allowed parameter
space to determine the structure of the Milky Way DM halo.

In order to implement these priors, we add a penalty factor to the
$\chi^2$ for each parameter (with Gaussian error), which takes into
account the external information.  The new $\chi^2$ reads
\begin{equation}
\chi^2_{\text{prior}} \left({\bf \theta}\right) =
\sum_{j=1}^{10} \sum_{i=1}^{20} \frac{\left( 
S_{ij}\left({\bf \theta}\right)  -  
S_{ij}^\text{th}\left({\bf \theta^0}\right) \right)^2} 
{(\sigma_{\rm stat})_{ij}^2 + (\sigma_{\rm sys})_{ij}^2} +
\left(\frac{\rho_\odot - \rho_\odot^0}{\sigma_{\rho_\odot}}\right)^2 +
\left(\frac{\log \langle \sigma v \rangle - \log \langle \sigma v
  \rangle^0}{\sigma_{\log \langle \sigma v \rangle}}\right)^2 ~,  
\label{Eq:chi2prior}
\end{equation}
where again, in $\log \langle \sigma v \rangle$, $\langle \sigma v
\rangle$ is expressed in cm$^3$/s.

In Figs.~\ref{Fig:80var} and~\ref{Fig:25var} we assume the same
parameters as in Figs.~\ref{Fig:80} and~\ref{Fig:25}, respectively,
but now we add the external information on $\rho_\odot$ and $\langle
\sigma v \rangle$ we have just discussed.  In both figures, the blue
(orange) regions correspond to the 68\% CL (90\% CL) contours for
2~dof.  In these figures, the case of our most optimistic set of
priors ($\sigma_{\log \langle \sigma v \rangle} = 0.15$ and
$\sigma_{\rho_\odot}= 0.1 \, \rho_\odot$) is represented by the dark 
regions, whereas the slightly less optimistic case ($\sigma_{\log
  \langle \sigma v \rangle} = 0.15$ and $\sigma_{\rho_\odot} =
0.5 \, \rho_\odot$) is depicted as the light regions. 

In both figures, we see that there is no further improvement on the
determination of the DM mass or the annihilation branching ratio, as
gamma-rays do not provide more stringent constraints on $\langle
\sigma v \rangle$ than what we assume to get from collider experiments
or DM direct detection searches.  Likewise, the ``data'' on gamma-rays
do not add further information on the local DM density.  Nevertheless,
for the case of $m_\chi = 80$~GeV (Fig.~\ref{Fig:80var}), the index
$\alpha$ could be constrained to lie in the range $\alpha \simeq
(0.13-0.22)$ and $\alpha \simeq (0.12-0.25)$ at 90\% CL (2~dof) and
for our most and less optimistic priors, respectively.  Although modest,
this could be an useful piece of information.  On the other hand,
$r_s$ could be only weakly constrained for the case with our most
optimistic priors, $r_s \simeq (8-60)~\text{kpc}$ at 90\% CL (2~dof).

Similar improvements are obtained for a lighter DM, as shown in
Fig.~\ref{Fig:25var} for $m_\chi = 25$~GeV.  Very small differences
are found between the results for these two DM masses.  The allowed
range for the index $\alpha$ would turn out to be $\alpha \simeq
(0.14-0.21)$ and $\alpha \simeq (0.13-0.23)$ at 90\% CL (2~dof) and
for our most and less optimistic priors, respectively.  On the other
hand, with our most optimistic priors, the allowed region for the
scale radius would be $r_s \simeq (8-56)~\text{kpc}$ at 90\% CL (2~dof).

\subsection{Comparison with numerical simulations}

The parametrization commonly used when studying the structural
properties of DM halos by means of numerical N-body simulations is the
NFW profile (Eq.~\eqref{Eq:NFW}).  Thus, in order to compare our
findings with their results, we consider this parametrization in the
next three figures.

\begin{figure}[t]
\begin{center}
\includegraphics[width=8.1cm]{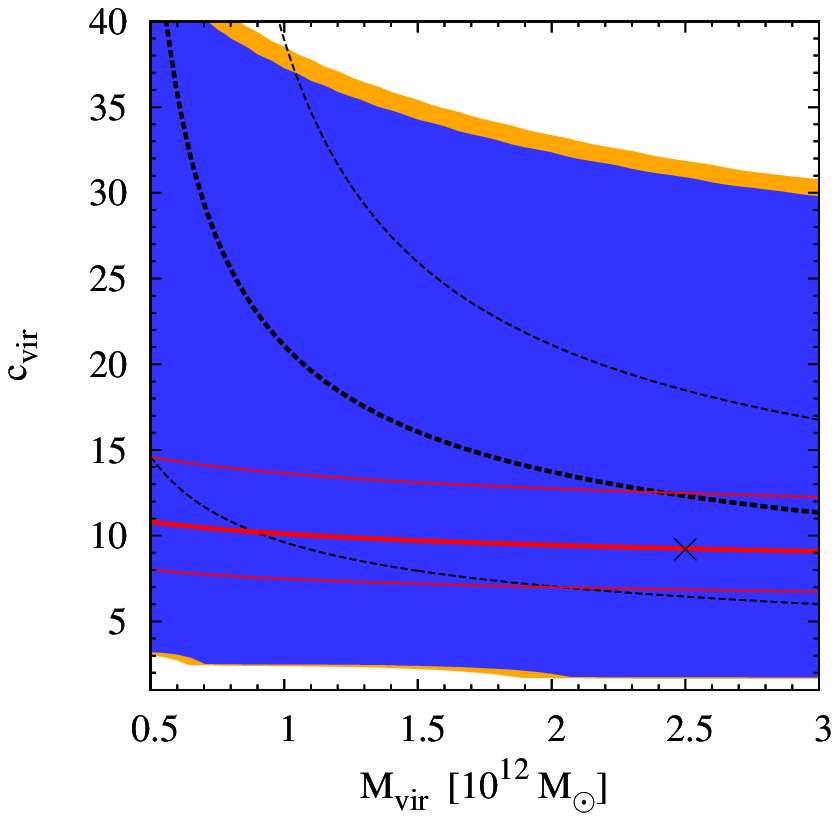}
\includegraphics[width=8.1cm]{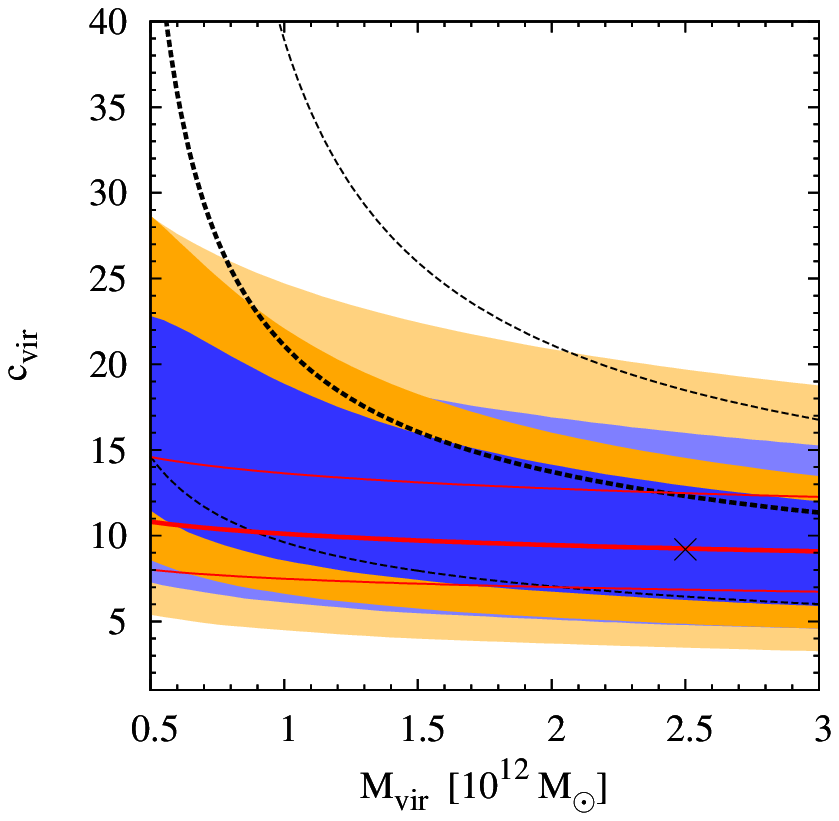}
\end{center}
\vspace{-0.8cm}
\caption{\sl \textbf{\textit{Fermi--LAT abilities to constrain the
      Milky Way DM density distribution after 5 years (keeping
      correlations with DM particle properties)}}.  We consider DM
    annihilation into a pure $b \bar b$ final state
    ($\text{BR}_{\tau(b)} = 0$), $\langle \sigma v \rangle = 3 \times
    10^{-26}$~cm$^3$/s and $m_\chi = 25$~GeV.  We assume a NFW
    profile, $\rho_\odot = 0.3$~GeV/cm$^3$, $R_\odot = 8.3$~kpc and
    the relation in Eq.~\eqref{Eq:cM} ($c_{\rm vir} = 9.2$, $r_s =
    38.6$~kpc and $M_{\rm vir} = 2.5 \times 10^{12} \, {\rm
      M}_\odot$).  The black crosses indicate the corresponding values
    in the $(M_{\rm vir}, c_{\rm vir})$ plane.  In both panels, blue
    (orange) regions represent the 68\% CL (90\% CL) contours for
    2~dof. The left panels show the case with no external priors and
    in the right panels two external priors are adopted: $\sigma_{\log
      \langle \sigma v \rangle} = 0.15$ and $\sigma_{\rho_\odot}= 0.1
    \, \rho_\odot$ for the dark regions and $\sigma_{\log \langle
      \sigma v \rangle} = 0.15$ and $\sigma_{\rho_\odot}= 0.5 \,
    \rho_\odot$ for the light contours.  Red solid lines indicate the
    mean concentration parameter obtained from numerical
    simulations~\cite{MunozCuartas:2010ig} and the $1\sigma$ scatter
    band ($\sigma_{\ln c_{\rm vir}} = 0.3$); black dashed lines
    indicate bottom-up $\rho_\odot = 0.2$~GeV/cm$^3$, $0.4$~GeV/cm$^3$
    and $0.6$~GeV/cm$^3$.}
\label{Fig:25cnfw}
\end{figure}

\begin{figure}[t]
\begin{center}
\includegraphics[width=8.1cm]{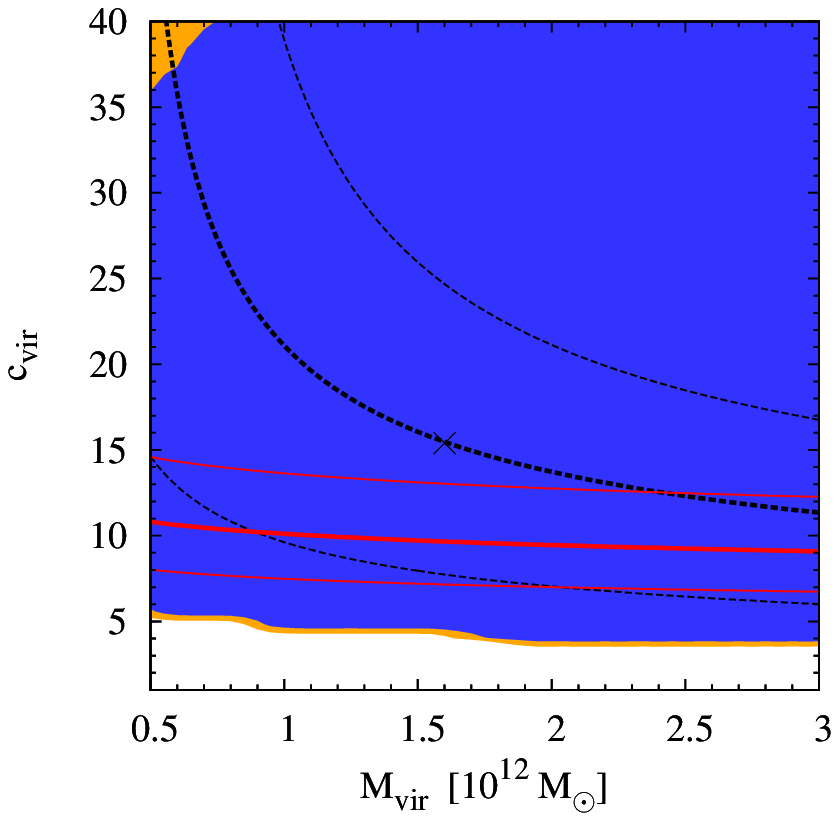}
\includegraphics[width=8.1cm]{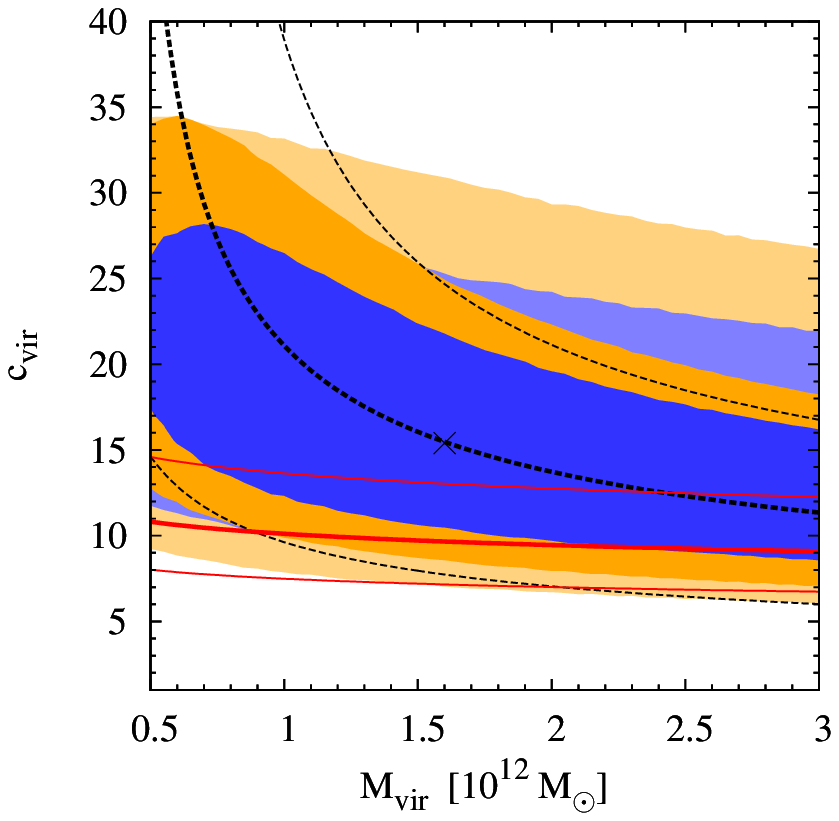}
\end{center}
\vspace{-0.8cm}
\caption{\sl \textbf{\textit{Fermi--LAT abilities to constrain the
      Milky Way DM density distribution after 5 years (keeping
      correlations with DM particle properties)}}.  Same as
  Fig.~\ref{Fig:25anfw}, but for $\rho_\odot = 0.4$~GeV/cm$^3$ and
  $r_s = 20$~kpc ($c_{\rm vir} = 15.2$ and $M_{\rm vir} = 1.6 \times
  10^{12} \, {\rm M}_\odot$).}
\label{Fig:25bnfw}
\end{figure}

\begin{figure}[t]
\begin{center}
\includegraphics[width=8.1cm]{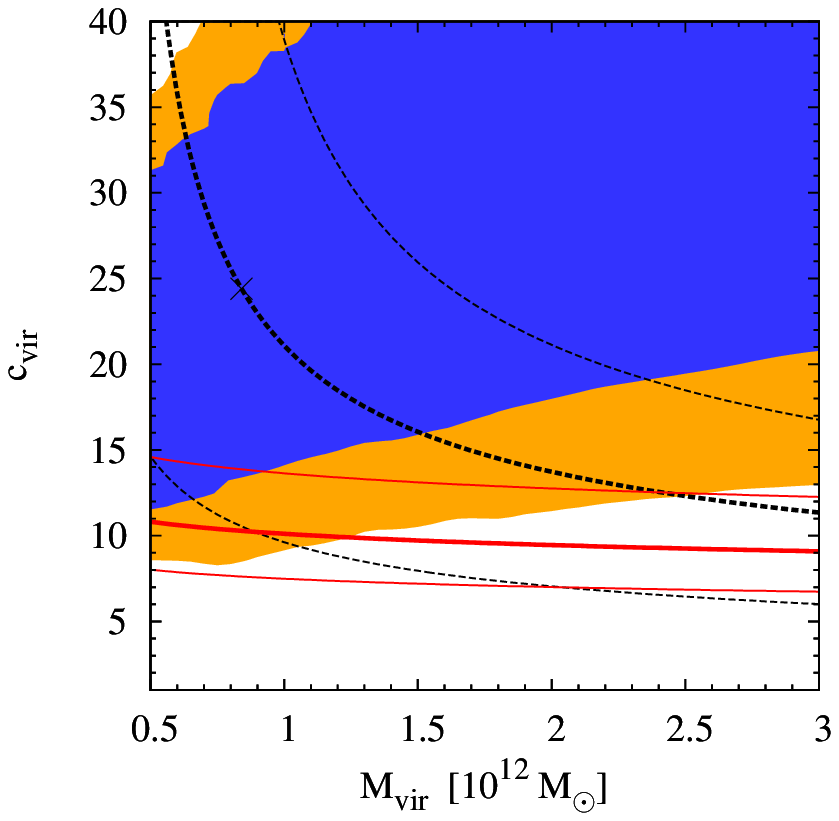}
\includegraphics[width=8.1cm]{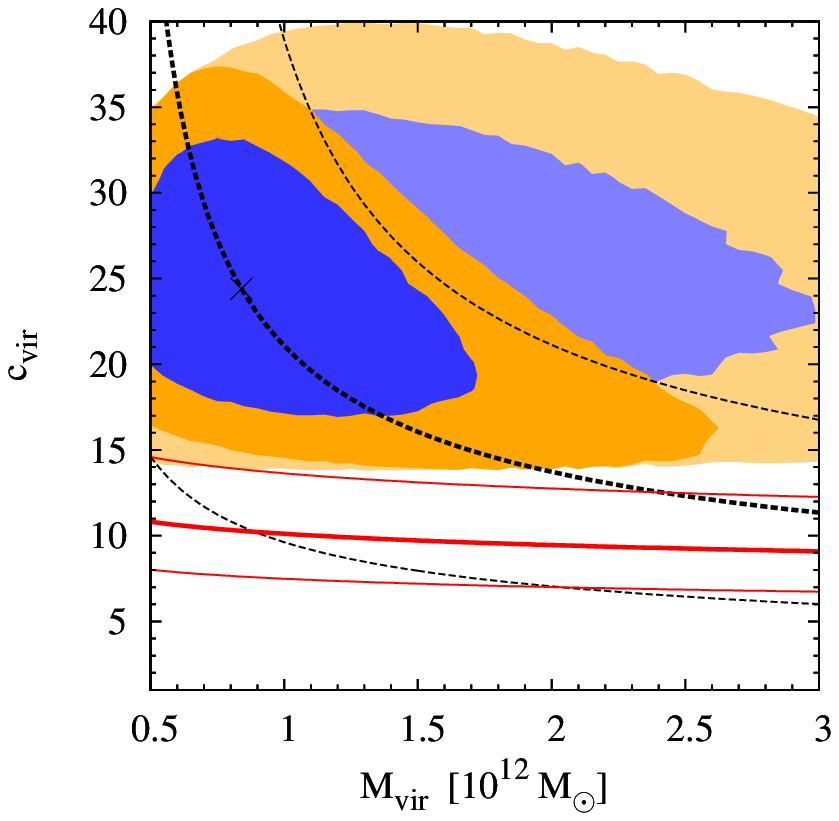}
\end{center}
\vspace{-0.8cm}
\caption{\sl \textbf{\textit{Fermi--LAT abilities to constrain the
      Milky Way DM density distribution after 5 years (keeping
      correlations with DM particle properties)}}.  Same as
  Fig.~\ref{Fig:25anfw}, but for $\rho_\odot = 0.4$~GeV/cm$^3$ and
  $M_{\rm dh} = 4.0 \times 10^{11} \, {\rm M}_\odot$ ($r_s = 10$~kpc,
  $c_{\rm vir} = 24.4$ and $M_{\rm vir} = 8.4 \times 10^{11} \, {\rm
    M}_\odot$).}
\label{Fig:25anfw}
\end{figure}

In Figs.~\ref{Fig:25cnfw}--\ref{Fig:25anfw}, we show the results for
different values of the parameters that describe the DM halo (we again
set $R_\odot = 8.3$~kpc), that is the local DM density and the scale
radius or equivalently, the virial mass and the concentration
parameter.  Indeed, for the sake of comparison with simulations, we
show the results in the plane $(M_{\rm vir}, c_{\rm
  vir})$~\footnote{We only show the range of $M_{\rm vir}$ currently
  favored.}, after marginalizing with respect to the three DM particle
parameters we study.  However, we do not show here the prospects to
constrain these parameters, as the results are very similar to those
obtained for the Einasto profile and depicted in the left panels of
Figs.~\ref{Fig:80}--\ref{Fig:25var}.  In the next figures, we assume
the same DM mass, $m_\chi = 25$~GeV and, as in
Figs.~\ref{Fig:80}--\ref{Fig:25var}, we consider DM annihilation into
a pure $b \bar b$ final state and an annihilation cross section (times
relative velocity) $\langle \sigma v \rangle = 3 \times
10^{-26}$~cm$^3$/s.

The left panels of Figs.~\ref{Fig:25cnfw}--\ref{Fig:25anfw} show the
capabilities of the {\it Fermi}--LAT experiment to reconstruct the
halo properties after 5 years.  For these panels no external priors are
assumed, analogously to Figs.~\ref{Fig:80} and~\ref{Fig:25}.  As in
the Einasto case, the blue and orange regions correspond to the 68\%
and 90\% CL contours for 2~dof, respectively.  Again, we see that
there are very small differences between these contours due to a very
steep $\chi^2$.

The right panels of Figs.~\ref{Fig:25cnfw}--\ref{Fig:25anfw} depict
the results in the case of adding external information on $\rho_\odot$
and $\langle \sigma v \rangle$, as in Figs.~\ref{Fig:80var}
and~\ref{Fig:25var}.  Likewise, the blue (orange) regions correspond
to the 68\% CL (90\% CL) contours for 2~dof.  In these panels, the
case of our most optimistic set of priors ($\sigma_{\log \langle
  \sigma v \rangle} = 0.15$ and $\sigma_{\rho_\odot}= 0.1 \,
\rho_\odot$) is also represented by the dark regions, whereas we show
the less optimistic case ($\sigma_{\log \langle \sigma v \rangle} =
0.15$ and $\sigma_{\rho_\odot} = 0.5 \, \rho_\odot$) as the light
regions. 

In all three figures, in both panels, the red solid curves represent
the mean value and $1\sigma$ scatter band of the concentration parameter
obtained in numerical simulations (see Eq.~\eqref{Eq:cM} and discussion
below).  The central black dashed line indicates our default value for
the local DM density, $\rho_\odot= 0.4$~GeV/cm$^3$, whereas the lower
black dashed line indicate $\rho_\odot = 0.2$~GeV/cm$^3$ and the upper
one $\rho_\odot = 0.6$~GeV/cm$^3$.

We first consider in Fig.~\ref{Fig:25cnfw} the case of $\rho_\odot =
0.3$~GeV/cm$^3$ and impose the relation $M_{\rm vir}-c_{\rm vir}$ for
the mean obtained in numerical simulations~\cite{MunozCuartas:2010ig}
(see Eq.~\eqref{Eq:cM}).  In this way, the two-parameter profile gets
reduced to a one-parameter parametrization.  Thus, a local DM
density~\footnote{Let us note that if $\rho_\odot = 0.4$~GeV/cm$^3$,
  then Eq.~\eqref{Eq:cM} implies $M_{\rm vir} = 6.1 \times 10^{12} \,
  {\rm M}_\odot$, which is probably too large.} $\rho_\odot =
0.3$~GeV/cm$^3$ and Eq.~\eqref{Eq:cM} imply $M_{\rm vir} = 2.5 \times
10^{12} \, {\rm M}_\odot$ and $c_{\rm vir} = 9.2$.  The scale radius
is predicted to be $r_s = 38.6$~kpc, which is much larger than the
value of 20~kpc commonly used in the literature of DM indirect
detection.  As we see from Fig.~\ref{Fig:25cnfw}, the gamma-ray data
alone would not be able to set any significant constraint on the DM
density profile.  Moreover, although adding external information on
$\rho_\odot$ and $\langle \sigma v \rangle$ would substantially reduce
the allowed parameter space, yet a large fraction would still be
unconstrained even in the most optimistic of our cases.  In particular,
no bounds on the virial mass could be set with this technique.

In general, the larger the scale radius (the smaller the concentration
parameter) the lower the potential signal from DM annihilations.  In
Fig.~\ref{Fig:25bnfw} we consider $\rho_\odot = 0.4$~GeV/cm$^3$ and 
$r_s =20$~kpc, which as mentioned above, is the common value used in
works related to indirect DM detection.  In this case, $M_{\rm vir} =
1.6 \times 10^{12} \, {\rm M}_\odot$ and $c_{\rm vir} = 15.2$.  Let us
note that this value for the scale radius implies a concentration
parameter much larger~\footnote{In the case of $\rho_\odot =
  0.3$~GeV/cm$^3$ and $r_s = 20$~kpc, one obtains $c_{\rm vir} =
  13.6$, which is still much larger than the predictions of the mean
  value in numerical simulations.} than what is obtained, for $M_{\rm
  vir} \sim 10^{12} \, {\rm M}_\odot$, for the mean in numerical
simulations, $c_{\rm vir} \sim 9-11$, and it is even away from the
$1\sigma$ scatter.  Thus, although certainly plausible, its choice is
not really justified by the value of the mean of the concentration
parameter obtained in numerical simulations.  It is interesting to
notice, though, that these parameters are very close to those obtained
in a very recent estimate~\cite{McMillan:2011wd}.  In the left panel of
Fig.~\ref{Fig:25bnfw}, we see that gamma-ray data alone would not be
able to offer any information on the DM density distribution in this
case either.  The addition of external constraints would allow to
substantially restrict the allowed parameter space, as can be seen
from the right panel.  However, only limited information could be
extracted, even in the case of our most optimistic priors.  At 90\%
CL (2~dof), the allowed parameter space would be compatible with the
mean obtained in numerical simulations.

On the other hand, in Fig.~\ref{Fig:25anfw} we use a recent estimate of
the dark halo mass within the innermost 60~kpc of the Milky
Way~\cite{Xue:2008se}, which predicts a lower mass than other works,
and we assume $M_{\rm dh} (60) = 4.0 \times 10^{11} \, {\rm M}_\odot$.
By considering a local DM density given by our default value,
$\rho_\odot = 0.4$~GeV/cm$^3$, this implies a scale radius much
smaller, $r_s = 10$~kpc, than what is usually assumed and a
concentration parameter correspondingly much larger, $c_{\rm vir} =
24.4$, than what is predicted from N-body simulations for halos with
virial masses about $10^{12} \, {\rm M}_\odot$ (cf.~Eq.~\eqref{Eq:cM}).
From the left panel, we see that a large part of the $1\sigma$ scatter
band obtained by numerical simulations would be disfavored by such a
measurement with only gamma-rays.  This trend is further accentuated
if external information on $\rho_\odot$ and $\langle \sigma v \rangle$
is available.  In that case, even with our less optimistic priors,
predictions of numerical simulations would be disfavored at more than
90\% CL, as can be seen from the right panel.  Hence, in this
situation a {\it Fermi}--LAT detection of DM would point out that a
revision of our current understanding of halo clustering and evolution
is mandatory, perhaps due to the influence of baryons, which are
usually not included in the simulations.

All in all, we see that numerical simulations tend to predict a mean 
concentration parameter which typically implies low statistics in a
gamma-ray experiment and thus, extracting information about the DM
density distribution would become a very challenging task in this
case.  In contrast, typical values for the DM local density imply large
values of the concentration parameter, which decrease with the virial
mass.  Hence, in order to avoid strong conflict with numerical
simulations, a virial mass in the high-mass side of the allowed
interval would be needed.  However, a recent determination of the DM
halo mass within the innermost 60~kpc~\cite{Xue:2008se} points to a
low virial mass, which would imply a large concentration parameter, in
clear conflict with numerical simulations (for commonly inferred
values of the DM local density), but with much more promising
prospects to constrain the halo density distribution with a gamma-ray
experiment like {\it Fermi}--LAT.

\section{Conclusions}
\label{conclusions}

During the last two decades, the progress in high resolution N-body
simulations has led to the discovery of a universal density profile for
CDM halos~\cite{Navarro:1995iw, Navarro:1996gj, Springel:2008cc,
  Navarro:2003ew, Graham:2005xx, Graham:2006ae, Graham:2006af,
  Navarro:2008kc} that can be parametrized in analytical forms
containing very few parameters.  However, these prescriptions only
represent the mean results for all simulated halos, but cannot
describe the large scatter that is found~\cite{Jing:1999km,
  Bullock:1999he, Wechsler:2001cs, AvilaReese:2005fm, Maccio':2006nu,
  Maccio':2008xb}.  Moreover, these simulations do not usually include
the effect of baryons, which could significantly change the
picture~\cite{Blumenthal:1985qy, Kazantzidis:2004vu, Gnedin:2004cx,
  Gustafsson:2006gr, Pedrosa:2009bt, Abadi:2009ve, Tissera:2009cm,
  Duffy:2010hf, Kazantzidis:2010jp}.  Hence, for the particular case
of our Milky Way, these mean values do not necessarily describe its CDM
halo.

This is a very important issue, as the predictions for the potential
signals for the astrophysical detection of DM critically depend on the
structure of the Milky Way halo.  Indeed, several approaches have been
proposed to determine DM particle properties by using future indirect
WIMP DM-induced signals of gamma-rays or neutrinos, direct detection
searches, collider information or their
combination~\cite{Dodelson:2007gd, Bernal:2008zk, Bernal:2008cu,   
  Jeltema:2008hf, Bernal:2010ip, Bernal:2010ti, PalomaresRuiz:2010pn,
  PalomaresRuiz:2010uu, Edsjo:1995zc, Cirelli:2005gh, Mena:2007ty,
  Agarwalla:2011yy, Das:2011yr, Lewin:1995rx, Primack:1988zm,
  Green:2007rb, Bertone:2007xj, Shan:2007vn, Drees:2008bv, Green:2008rd,
  Beltran:2008xg, Shan:2009ym, Strigari:2009zb, Peter:2009ak,
  Chou:2010qt, Shan:2010qv, Fox:2010bu, Kelso:2010sj,
  Bergstrom:2010gh, Pato:2010zk, Billard:2010jh, Shan:2011jz,
  Shan:2011ka, Drees:2000he, Polesello:2004qy, Battaglia:2004mp,
  Allanach:2004xn, Weiglein:2004hn, Birkedal:2005jq, Moroi:2005zx,
  Nojiri:2005ph, Baltz:2006fm, Arnowitt:2007nt, Cho:2007qv,
  Arnowitt:2008bz, Belanger:2008yc, Cho:2008tj, Baer:2009bu,
  Feldman:2011me, Bourjaily:2005ax, Altunkaynak:2008ry,
  Bertone:2010rv}.  All these measurements would be complementary and 
would constitute an important step toward the identification of the
particle nature of DM.  However, without a better knowledge of the
Milky Way DM density profile, large uncertainties would be present.
This, in the case of a positive signal, would make the task of
constraining DM properties a much more challenging one.

In the case of gamma-ray studies, although different profiles have
been considered in the literature, a single DM density profile is
commonly assumed to determine the capabilities of gamma-ray
experiments to extract some DM properties~\cite{Dodelson:2007gd,
  Bernal:2008zk, Bernal:2008cu, Jeltema:2008hf, Bernal:2010ip,
  Bernal:2010ti}, and thus in general, the obtained results are too
optimistic.  Here, we have studied the effect these astrophysical
uncertainties would have on the determination of some DM particle
properties, as annihilation cross section, mass and branching ratio
into dominant annihilation channels.  Conversely, gamma-ray searches
could also be used to learn about the structure of the Milky Way DM
halo, as a complementary tool to other type of observational data that
study the gravitational effect caused by the presence of DM.

In this work, we have studied the capabilities of the {\it Fermi}--LAT
instrument on board of the {\it Fermi} mission~\cite{Atwood:2009ez} to
tackle these issues and consider the potential gamma-ray signal in a
squared region with a side of $20^\circ$ around the GC ($|l|, |b| <
10^\circ$) after 5 years of data taking.  In order to model the
relevant gamma-ray backgrounds, we use the numerical code
GALPROP~\cite{Strong:1998pw} and the latest {\it Fermi}--LAT
observations~\cite{Collaboration:2010ru, Abdo:2010nz}.

In two introductory sections (Sections~\ref{gamma-rays}
and~\ref{MWprofile}) we review the main components of the gamma-ray
emission from DM annihilation in the GC, and critically discuss the
current knowledge of the parameters involved in the description of the
(smooth) Milky DM density profile.  Then, in Section~\ref{analysis} we
describe our modeling of the {\it Fermi}-LAT experiment and the
ingredients of the analysis we perform.  Finally, in
Section~\ref{results}, we present the results of our paper.  First, we
consider an Einasto profile and study the {\it Fermi}-LAT capabilities
to simultaneously constrain DM particle properties and the parameters
that describe the DM density profile.  We do so for two DM masses,
$m_\chi = 80$~GeV (Fig.~\ref{Fig:80}) and $m_\chi = 25$~GeV
(Fig.~\ref{Fig:25}).  These results could be compared to those of
Ref.~\cite{Bernal:2010ip}, where a particular DM density profile was
assumed to perform the fits, but with a single angular bin.  In
addition, we also consider the improvement in  these results when
external information on $\langle \sigma v \rangle$ and $\rho_\odot$ is
included (Figs.~\ref{Fig:80var} and~\ref{Fig:25var}).  In the last
part of this work, we focus on the determination of the parameters
that describe the DM density profile.  In order to compare our results
to the relations obtained in numerical simulations, we consider a NFW
profile, study different sets of values for the parameters and plot
the results in the $(M_{\rm vir}, c_{\rm vir})$ plane
(Figs.~\ref{Fig:25cnfw}--\ref{Fig:25anfw}).  The parameters and
relevant information for each of the figures are summarized in
Table~\ref{Tab:figs}.

In summary, and baring in mind the difficulties of all experiments
that aim to detect DM to distinguish that signal from any other possible
backgrounds, our study shows the {\it Fermi}--LAT capabilities to
simultaneously constrain DM particle properties and the Milky Way DM
density distribution.  Along these lines, we have tried to point out
some important issues that should be taken into account in indirect
searches when a potential DM signal is detected.

\section*{Acknowledgments}
We thank A.~Dutton, W.~Evans, J.~C.~Muñoz-Cuartas and M.~Wilkinson for
useful communications.  NB is supported by the DFG TRR33 `The Dark
Universe'.  SPR is partially supported by the Portuguese FCT through
CERN/FP/109305/2009 and CFTP-FCT UNIT 777, which are partially funded
through POCTI (FEDER), and by the Spanish Grant FPA2008-02878 of the
MICINN.

\small
\bibliographystyle{utphys}
\addcontentsline{toc}{section}{References}
\bibliography{biblio}

\end{document}